\documentclass[twocolumn,
    prc,nofootinbib,tightenlines,superscriptaddress,preprintnumbers]{revtex4-1}

\usepackage[utf8]{inputenc}
\usepackage{newtxtext,newtxmath}
\usepackage{bbold}
\usepackage{bm}

\usepackage{graphicx}
\usepackage[usenames,dvipsnames]{xcolor}
\usepackage{color}
\usepackage{tikz}

\usepackage[colorlinks=true,linkcolor=blue,urlcolor=blue,citecolor=blue]{hyperref}

\usepackage{amsmath}
\usepackage{slashed}

\usepackage[english]{babel}
\usepackage{titlesec}
\usepackage{microtype}
\usepackage{dcolumn}

\usepackage{blindtext}

\newcommand{\vect}[1]{\boldsymbol{#1}}

\titlespacing{\section}{4pt}{8pt plus 2pt minus 2pt}{4pt plus 2pt minus 2pt}
\titlespacing{\subsection}{0pt}{8pt plus 2pt minus 2pt}{4pt plus 2pt minus 2pt}
\titlespacing{\subsubsection}{0pt}{8pt plus 2pt minus 2pt}{0pt plus 2pt minus 2pt}

\graphicspath{{.}{figures/}}

\newcommand{\be}{\begin{equation}}
\newcommand{\ee}{\end{equation}}

\def\hs{\hspace}
\def\no{\nonumber}

\newcommand*\dd{\mathop{}\!\mathrm{d}}

\begin{document}
%\title{Symmetry energy of nuclear matter and isovector three-particle interactions}
%\title{Symmetry energy of nuclear matter and three-particle interactions in the Landau-Migdal theory} 
\title{Slope parameter of the symmetry energy \\
and the structure of three-particle interactions
in nuclear matter}
\author{Wolfgang Bentz}
\email[]{bentz@keyaki.cc.u-tokai.ac.jp}
\affiliation{Department of Physics, School of Science, Tokai University,
4-1-1 Kitakaname, Hiratsuka-shi, Kanagawa 259-1292, Japan}
%\affiliation{Radiation Laboratory, Nishina Center, RIKEN, Wako, Saitama 351-0198,
%Japan
\author{Ian C. Clo\"et}
\email[]{icloet@anl.gov}
\affiliation{Physics Division, Argonne National Laboratory, Argonne, Illinois 60439, USA}

\begin{abstract}
In the first part of this paper, we present a study of the symmetry energy ($a_s$) and its slope parameter ($L$) 
for nuclear matter in the framework of 
the Fermi liquid theory of Landau and Migdal. We derive an exact relation between $a_s$ and $L$, which involves the 
nucleon effective masses and
three-particle Landau-Migdal parameters. We present simple estimates which suggest that there are two
main mechanisms to explain the empirical values of $L$: The proton-neutron effective mass difference 
in isospin asymmetric matter and the $\ell=0$ moment of the isovector in-medium three-particle scattering amplitude. In the second part of this paper, we discuss the general structure of three-particle interactions in nuclear matter in the
framework of the Fermi liquid theory. The connections to the Bethe-Brueckner-Goldstone theory and
other approaches are also discussed. 
We show explicitly how the first few terms in the Faddeev series, together with medium induced three-particle interactions, 
emerge naturally in the Fermi liquid theory.  
        \newline\newline
    \noindent \textit{PhySH}: {
        Nuclear matter; 
        Nuclear forces;
        Nuclear many-body theory.
    }
\end{abstract}

\maketitle
%===============================================================================
%===============================================================================
\section{INTRODUCTION\label{sec:I}}
Among the basic physical quantities which determine the equation of state of nuclear systems, the symmetry energy ($a_s$) and its dependence on the baryon density ($\rho$) are receiving considerable attention recently because of their critical role in shaping the structure of nuclei and neutron stars~\cite{Horowitz:2000xj,Lattimer:2014sga, Baldo:2016jhp, Oertel:2016bki,Piekarewicz:2019ahf,Zhang:2020azr,Burgio:2021vgk}.  In medium to heavy nuclei with neutron excess,  the slope of the symmetry energy ($L \equiv 3 \rho \frac{\dd a_s}{\dd \rho}$),  which determines the associated symmetry pressure ($P_s = \frac{L}{3}\, \rho$), competes with  the surface tension to produce a neutron skin. A strong correlation between the skin thickness and the symmetry pressure has been reported~\cite{RocaMaza:2011pm,Baldo:2016jhp}. This subject  is currently under  experimental investigation for $^{208}$Pb and $^{48}$Ca nuclei at Jefferson 
Lab~\cite{Abrahamyan:2012gp,Tagami:2020shn,PREX:2021umo}. 

The symmetry pressure also works as a  restoring force in electric dipole oscillations, and analyses of experimental data have shown a strong correlation between the electric dipole polarizability and the neutron-skin thickness~\cite{Horowitz:2014bja,Baldo:2016jhp}. Also, other nuclear excitation modes of isovector character, like the quadrupole and spin-dipole resonances, appear to be sensitive to the density dependence of the symmetry energy~\cite{Colo:2013yta}.  In neutron stars the symmetry pressure competes with gravity to determine the radius of the star. A correlation between the neutron-skin thickness of $^{208}$Pb and the radius of a neutron star has been reported in recent analyses~\cite{Piekarewicz:2019ahf}. Finally, in the laboratory the density dependence of the symmetry energy can be probed in heavy-ion collisions by varying the energies and proton-neutron asymmetries of the colliding systems, and studying the isospin distributions among the reaction products~\cite{Baran:2004ih}.  Experimental programs in this direction are in progress or planned at various radioactive beam facilities.  A recent discussion of empirical values of $a_s$ and $L$, based on the different kinds of observations mentioned above, and their relation to the nucleon effective mass in-medium can be found in Ref.~\cite{Li:2018lpy}.

\looseness=-1
On the theoretical side, the most widely used frameworks to investigate the
density dependence of the symmetry energy are provided by extended parametrizations of 
Skyrme-type interactions~\cite{Goriely:2010bm,chen:2009wv,Zhang:2016aa,Somasundaram:2020chb}, 
relativistic mean field theory~\cite{Chen:2014sca,Dutra:2014qga}, chiral effective theories~\cite{Holt:2017uuq,Hammer:2019poc}, 
effective field theories based on low-momentum interactions~\cite{Bogner:2009bt,Drischler:2013iza}, and empirical 
parametrizations like metamodeling~\cite{Margueron:2017eqc}.  
In some of those approaches, effects of three-particle interactions
are incorporated by using density-dependent two-particle interactions, which is
of particular relevance for physical quantities related to third derivatives
of the energy density, like the skewness~\cite{Pearson:1991lsc} or the quantity $L$ mentioned above.
Many of these effective theories have their common roots in the more general framework of 
Landau's Fermi liquid theory~\cite{Landau:1956aa,Landau:1957aa,Landau:1959aa,Baym:2004aa}, and its extension
to nuclear systems by Migdal~\cite{Migdal:1967aa}. (For extensive reviews of the Landau-Migdal theory, see
for example Refs.~\cite{Speth:1977aa,Migdal:1990vm,Kamerdzhiev:2003rd}.) There is indeed a  close relationship between 
the Landau-Migdal approach and the Skyrme approach, as has been emphasized in Ref.~\cite{Speth:2014tja}.
The merit of the Fermi liquid theory is that it keeps model-dependent assumptions to an absolute
minimum, and exploits general symmetries like gauge invariance and Galilei invariance to derive
relations between the interaction parameters (Landau-Migdal parameters) and physical quantities 
which are in principle exact. In fact, it is now well known that the Fermi liquid theory
can be derived from the renormalization group~\cite{Shankar:1993pf}. The basic idea of this approach
is the concept of quasiparticles, which is well defined and useful near the Fermi surface.
For physical quantities which involve regions far away from the Fermi surface (for example the
bulk energy density or pressure of nuclear systems), more specific model assumptions must be made.

The purpose of the first part of this article is to derive an exact (model-independent) relation between the
symmetry energy and its slope parameter in the framework of the Fermi liquid theory of Landau and Migdal. 
We will show that this remarkably simple relation, which to the best of our knowledge has not
been presented so far in the literature, connects $a_s$ and $L$, at a certain
density, to the following physical quantities at the same density: the nucleon effective mass, the
slope of the proton-neutron effective mass difference arising from the isospin asymmetry, 
and two three-particle Landau-Migdal parameters, where only one of them (called $H_0'$ here) 
plays an important role. We will present semi-quantitative
discussions on each term in this relation, and compare the results with the empirical information. 
In view of the current interest in the symmetry energy and its slope parameter, and because of the
long history of studies on three-particle interactions in nuclear matter
~\cite{Bethe:1965zz,Day:1972zz,Day:1978tn,Day:1978zz,Day:1981zz,Day:1985jn},
we find it desirable to know such a model-independent relation based on first principles.
To derive this relation, we take the formalism of Ref.~\cite{Bentz:2019lqu},
where a similar relation between the skewness of nuclear matter ($J$) and three-particle interaction
parameters has been derived and discussed, and extend it to the isovector case.

The purpose of the second part of our work is to discuss the physical content of the in-medium three-particle amplitude,
the $\ell=0$ moments of which enter into the model-independent relations mentioned above. For this, we will extend the well known
discussions on the two-particle amplitude in the Fermi-liquid theory~\cite{Nozieres:1964aa,Klemt:1976zz,Poggioli:1976zz} to the three-particle case. 
Although the three-particle scattering amplitude 
in nuclear matter has been discussed in detail in the framework of the Bethe-Brueckner-Goldstone (BBG) theory
~\cite{Bethe:1965zz,Day:1972zz,Day:1978tn,Day:1978zz,Day:1981zz,Day:1985jn}, and the basic equations 
for the three-particle Green's function in nuclear systems are well known~\cite{Speth:1970aa,Ring:1974egq}, to the best of our
knowledge a discussion following the microscopic foundation of the Fermi-liquid theory has not yet been presented in the literature. 
For the purpose of deriving the basic formulas for the three-particle amplitude
in this framework, we will limit ourselves to the case of symmetric nuclear matter. We will
discuss how far the structure of the three-particle amplitude can be specified by using only its
definition, and illustrate how further assumptions, similar to
the ones used in the BBG theory, can be used to derive more detailed expressions.
Among those expressions, we will recover terms of the familiar Faddeev series
~\cite{Faddeev:1960su}, and also
terms of four-body nature which arise from the interaction of the three given particles 
with the Fermi sea.

The BBG theory mentioned above, which is based on the hole-line expansion of the energy density~\cite{Mahaux:1979eor}, has been extensively used recently by using modern two- and three-nucleon potentials~\cite{Song:1998zz,Zuo:2002sg,Lu:2017nbi,Lu:2018dmn}. It is still one of the most important methods, commonly called {\it ab initio}
microscopic methods, to determine the equation of state of nuclear systems. Other {\it ab initio} microscopic 
methods are based on the variational method~\cite{Pandharipande:1979bv,Hagen:2013yba}, the self-consistent Green's function method~\cite{Dickhoff:2004xx,Carbone:2013eqa},
and Quantum Monte Carlo methods~\cite{Carlson:2003wm,Gandolfi:2009fj}. All these important theoretical tools aim to improve the 
quantitative
understanding of saturation properties, effects of neutron excess related to the symmetry energy and its slope, 
and the equation of state at high baryon
densities. As we explained already above, the aim of our present work is different: First, we wish to exploit 
the predictive power of the
Fermi liquid theory to relate three-particle interaction parameters to physical quantities of nuclear matter connected to
the symmetry energy. 
Second, we wish to elucidate the structure of the three-particle in-medium scattering amplitude
as it follows from its general definition, and illustrate the relation to the BBG theory by making further
model dependent assumptions.

The layout of the paper is as follows: In Sec.~\ref{sec:II} we use the Fermi liquid theory of Landau and Migdal to derive our relation between the slope parameter of the symmetry energy
and the  three-particle interaction parameters, and present a semi-quantitive discussion of this relation
in connection to empirical values. In Sec.~\ref{sec:III} we discuss the physical content of the three-particle 
scattering amplitude in the Fermi liquid theory, and make connection to the BBG theory.    
In Sec.~\ref{sec:IV} we summarize our results, and further comment on the relation between our approach
and other methods mentioned above. 
App.~\ref{app:A} is devoted to a detailed discussion of Galilei invariance relations for
isospin asymmetric nuclear systems, and in App.~\ref{app:B} we prove several relations which are used in Sec.~\ref{sec:III}.

%==========================================================================    
%=========================================================================
\section{SYMMETRY ENERGY AND ITS SLOPE PARAMETER IN THE LANDAU-MIGDAL THEORY\label{sec:II}}
The aim of this Section is first to use the Landau-Migdal theory of nuclear matter to derive an exact relation between the
symmetry energy and its slope parameter in terms of the nucleon effective mass and three-particle
Landau-Migdal parameters.
Second, we wish to present a semi-quantitative discussion of this relation by approximating the three-particle
interaction parameters by simple expressions which follow from the driving term of the Faddeev equation, and compare
the results with empirical values.

\subsection{Theoretical framework\label{sec:IIa}}
In order to  discuss the density dependence of the symmetry energy of nuclear matter in a general framework,  
we extend the basic formula of the Fermi liquid theory~\cite{Negele:1988aa} for spin-independent but isospin dependent variations of the energy 
density $E$ to include the third order term:
\begin{align}
\delta E&(\{\rho\}) = 2 \int \frac{{\rm d}^3 k}{(2 \pi)^3}\ \varepsilon^{(\tau)}({\vect{k}}; \{\rho\}) \, \delta n^{(\tau)}_{\vect{k}}
\nonumber \\ 
&+ \frac{1}{2}  \left( \prod_{i=1}^2 \, 2 \int \frac{{\rm d}^3 k_i}{(2 \pi)^3} \, 
\delta n^{(\tau_i)}_{{\vect{k}}_i} \right) \, f^{(\tau_1 \tau_2)} (\vect{k}_1, \vect{k}_2; \{\rho\})  \nonumber \\  
&+ \frac{1}{6}  \left( \prod_{i=1}^3 \, 2 \int \frac{{\rm d}^3 k_i}{(2 \pi)^3} \, 
\delta n^{(\tau_i)}_{{\vect{k}}_i} \right) \, h^{(\tau_1 \tau_2 \tau_3)}(\vect{k}_1, \vect{k}_2; \vect{k}_3 ; \{\rho\}).
\label{vare}
\end{align}
Here $\{\rho\} \equiv \{\rho^{(p)}, \rho^{(n)}\}$ represents an arbitrary set of proton and neutron 
background densities. 
%and $\{\rho_0\} \equiv \{\rho_0^{(p)}, \rho_0^{(n)}\}$ are reference densities. 
The superscript $(\tau)$ distinguishes between
protons ($\tau=p$) and neutrons ($\tau=n$), and summations over all $\tau$'s are implied. 
The energy of a quasiparticle with momentum $\vect{k}$ is denoted as
$\varepsilon^{(\tau)}({\vect{k}};\{\rho\})$,
$f^{(\tau_1 \tau_2)} (\vect{k}_1, \vect{k}_2; \{\rho \})$ is the spin-averaged  
forward scattering amplitude of two quasiparticles with momenta $\vect{k}_1, \vect{k}_2$,  
%at background density $\{\rho\}$; 
and $h^{(\tau_1 \tau_2 \tau_3)}(\vect{k}_1, \vect{k}_2; \vect{k}_3 ; \{\rho\})$ is the corresponding three-particle forward 
scattering amplitude. 
The functions $f$ and $h$ are symmetric with respect to simultaneous interchanges of the momentum and isospin  
variables, and can be represented by a set of connected diagrams with four and six external nucleon
lines, respectively. Density variations and quasiparticle energies which are independent of the direction of $\vect{k}$ 
will be denoted as $\delta n^{(\tau)}_k$ and $\varepsilon^{(\tau)}(k; \{\rho \})$.

The form of $\delta n^{(\tau)}_{{k}}$, corresponding to an isospin dependent change of the Fermi momentum\footnote{In this paper $p$ denotes a Fermi momentum, 
i.e., $p^{(p)}$ and $p^{(n)}$ are the Fermi momenta of protons and
neutrons, and $p$ is the Fermi momentum for the isospin symmetric case. The relation to the densities is given by
$\rho^{(\tau)} = \frac{p^{(\tau)3}}{3 \pi^2}$ and $\rho = \frac{2 p^3}{3 \pi^2}$.
Quasiparticle energies, effective masses, and scattering amplitudes without arguments are defined at their respective Fermi
surfaces,  e.g., $\varepsilon^{(\tau)} \equiv \varepsilon^{(\tau)}(k=p^{(\tau)}; \{\rho\})$, 
$\, M^{*(\tau)} \equiv M^{* (\tau)}(k=p^{(\tau)}; \{\rho\})$,  
$\, f_0^{(\tau_1 \tau_2)} \equiv f_0^{(\tau_1 \tau_2)}(k_1=p^{(\tau_1)}, k_2=p^{(\tau_2)}; \{\rho\})$, etc. 
Quantities without isospin variables, or with a single symbol for the background density ($\rho$),
refer to the limit of isospin symmetry ($\rho^{(3)}=0$).}
by $\delta p^{(\tau)}$, is given to first order by
\begin{align}
\delta n^{(\tau)}_{k} = 
\delta p^{(\tau)} \cdot \delta(p^{(\tau)} - k) = \frac{\pi^2}{p^{(\tau)2}} \, \delta \rho^{(\tau)} \cdot  \delta(p^{(\tau)} - k) \,.  
\label{dn}
\end{align}
The first order variation of $E$ is then given by
%\footnote{The symbol $\frac{\delta}{\delta \rho^{\tau}}$ denotes the derivative w.r.t. only the background densities,
%and $\frac{\partial}{\partial \rho^{\tau}} = \frac{\pi^2}{p^{(\tau)2}} \frac{\partial}{\partial p^{(\tau)}} + 
%\frac{\delta}{\delta \rho^{\tau}}$ denotes the derivative w.r.t. the density $\rho^{\tau}$. Quasiparticle energies and
%scattering amplitudes, for which the external momenta are equal to the respective Fermi momenta, will be denoted
%without arguments, e.g., $\varepsilon^{(p)} \equiv \varepsilon^{(p)}(k=p^{(p)}; \{\rho\})$, 
%$f_0^{(pn)} \equiv f_0^{(pn)}(k_1=p^{(p)}, k_2=p^{(n)}; \{\rho\})$, etc.}
%
\begin{align}
\frac{\delta E(\{\rho\})}{\delta \rho^{(\tau)}} = \varepsilon^{(\tau)}(p^{(\tau)}; \{\rho\})  \equiv \varepsilon^{(\tau)}\,.   
\label{eps}
\end{align}
Here and in the following, the symbol ${\displaystyle \frac{\delta}{\delta \rho^{(\tau)}}}$ 
denotes the derivative w.r.t. the background densities, keeping external momenta (if any) fixed, while
${\displaystyle \frac{\partial}{\partial \rho^{(\tau)}}}$ includes also the derivative
w.r.t. external momentum variables, if those are equal to the Fermi momentum $p^{(\tau)}$.

It is convenient to express Eq.~(\ref{eps}) and the following relations by using the
sum and difference of proton and neutron densities:
\begin{align}
\rho = \rho^{(p)} + \rho^{(n)} \,, \,\,\,\,\,\,\,\,\,\,\,\,\,\,\,\,\,\,\,
\rho^{(3)} = \rho^{(p)}  - \rho^{(n)} \,.
\label{rpn} 
\end{align}
Then Eq.~(\ref{eps}) can be written as 
\begin{align}
\frac{\delta E(\{\rho\})}{\delta \rho} &= \frac{1}{2} 
\left(\varepsilon^{(p)} + \varepsilon^{(n)} \right) \,, 
\label{epss} \\ 
\frac{\delta E(\{\rho\})}{\delta \rho^{(3)}} &= \frac{1}{2} 
\left(\varepsilon^{(p)}  - \varepsilon^{(n)} \right) \,.
\label{epsv}
\end{align}
The first order variation of the quasiparticle energy $\varepsilon^{(\tau)}(k; \{\rho\})$ 
w.r.t. the background densities is given by
\begin{align}
\frac{\delta \varepsilon^{(\tau_1)}(k_1; \{\rho\})}{\delta \rho^{(\tau_2)}} = f_0^{(\tau_1 \tau_2)}(k_1, k_2=p^{(\tau_2)}; \{\rho\}) \,.
\label{deps}
\end{align}
The $\ell=0,1$ moments of the forward scattering amplitude are defined as usual by
\begin{align}
\frac{1}{2 \ell +1} \, &f_{\ell}^{(\tau_1 \tau_2)}(k_1, k_2; \{\rho\}) \nonumber \\
&= \int \frac{{\rm d}\Omega_2}{4 \pi} 
\left({\vect{\hat k}}_1 \cdot {\vect{\hat k}}_2 \right)^{\ell} \, f^{(\tau_1 \tau_2)}(\vect{k}_1, \vect{k}_2; \{\rho\}) \,.
\label{f01}
\end{align}
We can use Eq.~(\ref{deps}) to extract information on the density dependence of the effective masses
of protons and neutrons, 
which are defined as usual in terms of the quasiparticle velocity by
${\displaystyle \frac{\partial \varepsilon^{(\tau)}(k; \{\rho\})}{\partial k}
\equiv \frac{k}{M^{*(\tau)}(k; \{\rho\})}}$. For this, we take the partial derivative of Eq.~(\ref{deps})
w.r.t. $k_1$ and then set $k_1=p^{(\tau_1)}$. This gives
\begin{align}
\frac{\delta M^{*(\tau_1)}}{\delta \rho^{(\tau_2)}} = - \frac{M^{*(\tau_1)2}}{p^{(\tau_1)}} \,
\frac{\partial f_0^{(\tau_1 \tau_2)}}{\partial p^{(\tau_1)}} \,.   
\label{relb}
\end{align}
%where quantities without arguments refer to the Fermi surface, i.e., $M^{*(\tau)} \equiv M^{*(\tau)}(p^{(\tau)}; \{\rho\})$,  
%$f_0^{(\tau_1 \tau_2)} \equiv f_0^{(\tau_1 \tau_2)}(p^{(\tau_1)}, p^{(\tau_2)}; \{\rho\})$.
In the isospin symmetric limit ($\rho^{(3)} \rightarrow 0$) we obtain from Eq.~(\ref{relb})
\footnote{Hereafter, in the rest of this
paper (including App.~\ref{app:A}), all derivatives w.r.t. $\rho^{(3)}$ are defined at $\rho^{(3)}=0$,
although this is not indicated explicitly in order to simplify the notation.}
\begin{align}
\frac{\delta M^{*}}{\delta \rho} &= - \frac{M^{*2}}{2 p} \, \frac{\partial f_0}{\partial p} \,,  
\label{bis} \\ 
\frac{\delta M^{*(p)}}{\delta \rho^{(3)}}
&= - \frac{\delta M^{*(n)}}{\delta \rho^{(3)}}
= - \frac{M^{*2}}{2 p} \, \frac{\partial f_0'}{\partial p} \,.  
\label{biv}
\end{align} 
Here we defined the functions~\cite{Migdal:1967aa}  
\begin{align}
f_0 = \frac{1}{2} \left (f_0^{(pp)} + f_0^{(pn)} \right), \,\,\,\,\,\,
f'_0 = \frac{1}{2} \left (f_0^{(pp)} - f_0^{(pn)} \right),
\label{f0}
\end{align}
in the isospin symmetric limit. 
The partial derivative of $f_0 \equiv f_0(p, p; \rho)$ 
and $f'_0 \equiv f'_0(p, p; \rho)$ w.r.t. the Fermi momentum $p$ by definition acts 
on both momentum variables, e.g., for $f_0$,
\begin{align}
\frac{\partial f_0}{\partial p} \equiv \left[ \left(\frac{\partial}{\partial k_1} + 
\frac{\partial}{\partial k_2}\right) \, f_0(k_1, k_2; \rho) \right]_{k_1 = k_2 = p},
\label{ddef}
\end{align}
and because of the symmetry of the scattering amplitude this is the same as the derivative 
w.r.t. only one momentum variable, multiplied by 2.   
Eqs.~(\ref{bis}) and (\ref{biv}) lead to the following
expressions for the ``total'' derivatives of the effective masses w.r.t. the densities:  
\begin{align}
\left(\frac{{\partial} M^{*}}{{\partial } \rho}\right) 
&= - \frac{M^{*2}}{2 p} \, \frac{\partial f_0}{\partial p} 
+ \frac{\pi^2}{2 p^2} \frac{\partial M^*}{\partial p}\,,   \label{c1} \\
\left(\frac{{\partial} M^{*(p)}}{{\partial} \rho^{(3)}}\right)
&= - \left(\frac{{\partial} M^{*(n)}}{{\partial} \rho^{(3)}}\right)
\nonumber \\
&= - \frac{M^{*2}}{2 p} \, \frac{\partial f_0'}{\partial p} 
+ \frac{\pi^2}{2 p^2} \frac{\partial M^*}{\partial p} \,.  \label{c2} 
\end{align}
We will make use of these relations in later developments.

From Eq.~(\ref{deps}) we obtain the following relation for the
derivatives of the Fermi energies $\varepsilon^{(\tau)}$:
\begin{align}
\frac{\partial \varepsilon^{(\tau_1)}}{\partial \rho^{(\tau_2)}}
= \delta_{\tau_1, \tau_2} \, \frac{\pi^2}{M^{*(\tau_1)} \, p^{(\tau_1)}} \,  
+ f^{(\tau_1 \tau_2)}_0 \,.
\label{deps1}
\end{align}  
This relation, together with Eqs.~(\ref{epss}) and (\ref{epsv}), leads to the following well known 
expressions for the second derivatives of the energy density in the isospin symmetric limit
($\rho^{(3)} \rightarrow 0$):
\begin{align}
\frac{\partial^2 E}{\partial \rho^2} &= \frac{\pi^2}{2 p \, M^*}  + f_0 \equiv  
\frac{\pi^2}{2 p \, M^*} \left( 1 + F_0 \right)  \equiv \frac{K}{9 \rho} \,, \label{ds} \\
\frac{\partial^2 E}{\partial \rho^{(3)2}} &= \frac{\pi^2}{2 p \, M^*}  + f'_0 \equiv  
\frac{\pi^2}{2 p \, M^*} \left( 1 + F'_0 \right)  \equiv \frac{2 a_s}{\rho} \,. \label{dv} 
\end{align}
Here we defined the dimensionless Landau-Migdal parameters $F_0$ and $F_0'$, the 
incompressibility $K$, and the symmetry energy $a_s$ in the usual way~\cite{Negele:1988aa}.

The derivative of the symmetry energy $a_s$ w.r.t. the density is obtained from
the definition, given in Eq.~(\ref{dv}), as
\begin{align}
\frac{{\rm d} a_s}{{\rm d} \rho} = \frac{\pi^2}{6 p M^*} + \frac{1}{2} \, f_0' 
- \frac{p^2}{6 M^{*2}} \, \frac{{\rm d} M^*}{{\rm d} \rho} + \frac{p^3}{3 \pi^2} \,  
\frac{{\rm d} f_0'}{{\rm d} \rho} \,.
\label{asp}
\end{align}
In order to specify the last term in this relation, we note that in the isospin symmetric limit the derivative of
$f_0'= \left(f_0^{(pp)} - f_0^{(pn)}\right)/2$ w.r.t. the background density is obtained from Eqs.~(\ref{vare}) and (\ref{dn}) as
\begin{align}
\frac{\delta f_0'}{\delta \rho} &= \frac{1}{4} \sum_{\tau} \left(h_0^{(pp\tau)} - h_0^{(pn\tau)} \right)
= \frac{1}{4} \left( h_0^{(ppp)} - h_0^{(ppn)} \right) \equiv \, h_0' \,.
\label{h0p}    
\end{align}
Here we define the $\ell=0,1$ moments of the three-particle amplitude as
\begin{align}
&\frac{1}{2 \ell +1} \, h_{\ell}^{(\tau_1 \tau_2 \tau_3)}(k_1, k_2, k_3; \rho) \nonumber \\  
&= \int \frac{{\rm d} \Omega_2}{4 \pi}\, \int \frac{{\rm d} \Omega_3}{4 \pi} \,
\left( \hat{\vect{k}}_1 \cdot \hat{\vect{k}}_2 \right)^{\ell}  
\, h^{(\tau_1 \tau_2 \tau_3)}(\vect{k}_1,\vect{k}_2, \vect{k}_3; \rho)  \,. 
\label{old4}
\end{align}
The first equality in Eq.~(\ref{h0p}) follows from the general definition of the three-particle amplitude
according to Eq.~(\ref{vare}), 
and the second equality holds in the isospin symmetric limit, where the interchange $p \leftrightarrow n$ is possible, and the case of $\ell=0$ in Eq.~(\ref{old4}).

For later comparison we note that the isoscalar counterpart of Eq.~(\ref{h0p}) is given by
\begin{align}
\frac{\delta f_0}{\delta \rho} &= \frac{1}{4} \sum_{\tau} \left(h_0^{(pp\tau)} + h_0^{(pn\tau)} \right) \no \\
&= \frac{1}{4} \left( h_0^{(ppp)} + 3 h_0^{(ppn)} \right) \equiv \, h_0 \,.
\label{h0}   
\end{align}
By using Eqs.~(\ref{h0p}) and (\ref{c2}) we can express the derivative of $f_0'$ w.r.t. the density in the following way:
\begin{align}
\frac{{\rm d} f_0'}{{\rm d} \rho} &= h_0' + \frac{\pi^2}{2 p^2} \, \frac{\partial f_0'}{\partial p}
\nonumber \\
&=  h_0' + \frac{\pi^2}{2 p^3} \left(- \frac{p^2}{M^{*2}} \frac{\partial \Delta M^*}{\partial \rho^{(3)}}  
+ \frac{\pi^2}{M^{*2}} \frac{\partial M^*}{\partial p} \right) \,, 
\label{der}
\end{align}
where $\Delta M^* \equiv M^{*(p)} - M^{*(n)}$ denotes the difference of proton and neutron effective
masses arising from the isospin asymmetry to first order in $\rho^{(3)}$.

Eq.~(\ref{der}) summarizes the result for the last term in Eq.~(\ref{asp}). For the third term
of Eq.~(\ref{asp}), we can make use of Eq.~(\ref{c1}) and the following relation, which follows from 
Galilei invariance (see Eqs.~(18) and (19) of Ref.~\cite{Bentz:2019lqu}):
\begin{align}
- p \frac{\partial f_0}{\partial p} &= 
\frac{1}{3} \,  p \frac{\partial f_1}{\partial p} + \frac{4}{3} \, f_1 +  
\frac{4 p^3}{3 \pi^2} \, h_1  \label{old5a} \\ 
&= - \frac{3 \pi^2}{p} \frac{M-M^*}{M M^*} + \frac{\pi^2}{M^{*2}} 
\frac{\partial M^*}{\partial p} + \frac{4 p^3}{3 \pi^2} \, h_1 \,.
\label{old5}
\end{align}
Here $h_1$ is the $\ell=1$ moment of the isoscalar three-particle amplitude given in Eq.~(\ref{old4}), that is, in the isospin symmetric limit 
${\displaystyle h_1 = \frac{1}{4} \left( h_1^{(ppp)} + 3 h_1^{(ppn)} \right)}$, which agrees with the
isospin average considered in Ref.~\cite{Bentz:2019lqu} from the outset.

We now insert all results into Eq.~(\ref{asp}) to obtain
\begin{align}
\frac{{\rm d} a_s}{{\rm d} \rho} &= 
\frac{\pi^2}{4 p M^*} \left[ \frac{2}{3} + \frac{M-M^*}{M} + F_0' - \frac{M}{M^*} \mu   
+ H_0' - \frac{1}{3} H_1 \right]\,.
\label{r2}
\end{align} 
Here we defined the dimensionless three-particle interaction parameters by~\cite{Bentz:2019lqu}
\begin{align}
H_{\ell} = \frac{4 p^4 M^*}{3 \pi^4} \, h_{\ell} \,, \,\,\,\,\,\,
H'_{\ell} = \frac{4 p^4 M^*}{3 \pi^4} \, h'_{\ell}  \,,      
\label{dim1}
\end{align}
and introduced the density dependent quantity $\mu(\rho)$ according to
\begin{align}
\mu \equiv \rho \, \frac{\partial}{\partial \rho^{(3)}} \left(\frac{\Delta M^*}{M} \right) \,.   
\label{mu}
\end{align}   
One should note that the terms ${\displaystyle \propto \frac{\partial M^*}{\partial p}}$ canceled in
the result given in Eq.~(\ref{r2}).

Eq.~(\ref{r2}) can be expressed in terms of the slope parameter $L$ of the symmetry
energy, which is defined as usual by ${\displaystyle L = 3 \rho \frac{{\rm d} a_s}{{\rm d} \rho}}$.
Using also ${\displaystyle a_s = \frac{p^2}{6 M^*} \left(1 + F_0'\right)}$ according to Eq.~(\ref{dv}), 
we obtain finally 
\begin{align}
L &= 3 a_s  - \frac{p^2}{2M} \nonumber \\
&\times \left[\left(1 - \frac{2}{3} \frac{M}{M^*} \right) + \mu \, \left(\frac{M}{M^*}\right)^2 
- \frac{M}{M^*} \left(H_0' - \frac{1}{3} H_1 \right)\right]\,.  
\label{result}
\end{align}
One should note that Eq.~(\ref{result}) is an exact relation.

%===============================================================================
%===============================================================================
\subsection{Empirical values and semi-quantitative discussions\label{sec:IIb}}
As discussed in Sec.~\ref{sec:I}, there have been many investigations in the literature on how to extract empirical 
values of the symmetry energy and the slope parameter at normal nuclear matter density ($\rho_0 = 0.155$ fm$^{-3}$) as well
as subnormal and higher densities. Although there are correlations between these two quantities, for definiteness we refer here
to the experimental data summarized in Fig.~20 of
Ref.~\cite{Li:2018lpy} for each quantity separately, from which the following fiducial values have been extracted:
\begin{align}
a_s = 31.6 \pm 2.66~{\rm MeV}\,,
\,\,\,\,\,\,\,\,\,\,\,\,L = 59 \pm 16~{\rm MeV}\,. 
\label{emp}
\end{align}
We mention that these values are consistent with most of the other analyses mentioned in Sec.~\ref{sec:I}.
In particular, they are consistent with the more stringent constraint $a_s - \frac{L}{9} \simeq (25 - 26)$ MeV reported in Ref.~\cite{Chen:2014sca} as well as in various previous references~\cite{Brown:2000pd,Furnstahl:2001un,Brown:2013mga}, 
and also encompass the values $a_s = 31 \pm 2$ MeV, $L = 55 \pm 12$ MeV
reported very recently in Ref.~\cite{Tagami:2020shn}. 

In the analysis of Ref.~\cite{Li:2018lpy}, the following empirical values of the quantity $\mu$, defined by
Eq.~(\ref{mu}), have also been reported:
\begin{align}
\mu = 0.27 \pm 0.25  \,. 
\label{mu1}
\end{align}
For simplicity, in the following discussion we will assume that $0 < \mu < 0.5$. 
Concerning the nucleon effective mass $M^*$, we will consider the same conservative limits
as in our previous work~\cite{Bentz:2019lqu}:
\begin{align}
0.7 \, < \, \frac{M^*}{M} \, < \, 1.0 \,,   
\label{mst}
\end{align}
which encompasses the values reported by intensive investigations during the last decades~\cite{Mahaux:1985zz,Blaizot:1981zz,vanDalen:2005sk,Li:2018lpy}.

\begin{figure}
\begin{center}
\includegraphics[scale=0.4,angle=90]{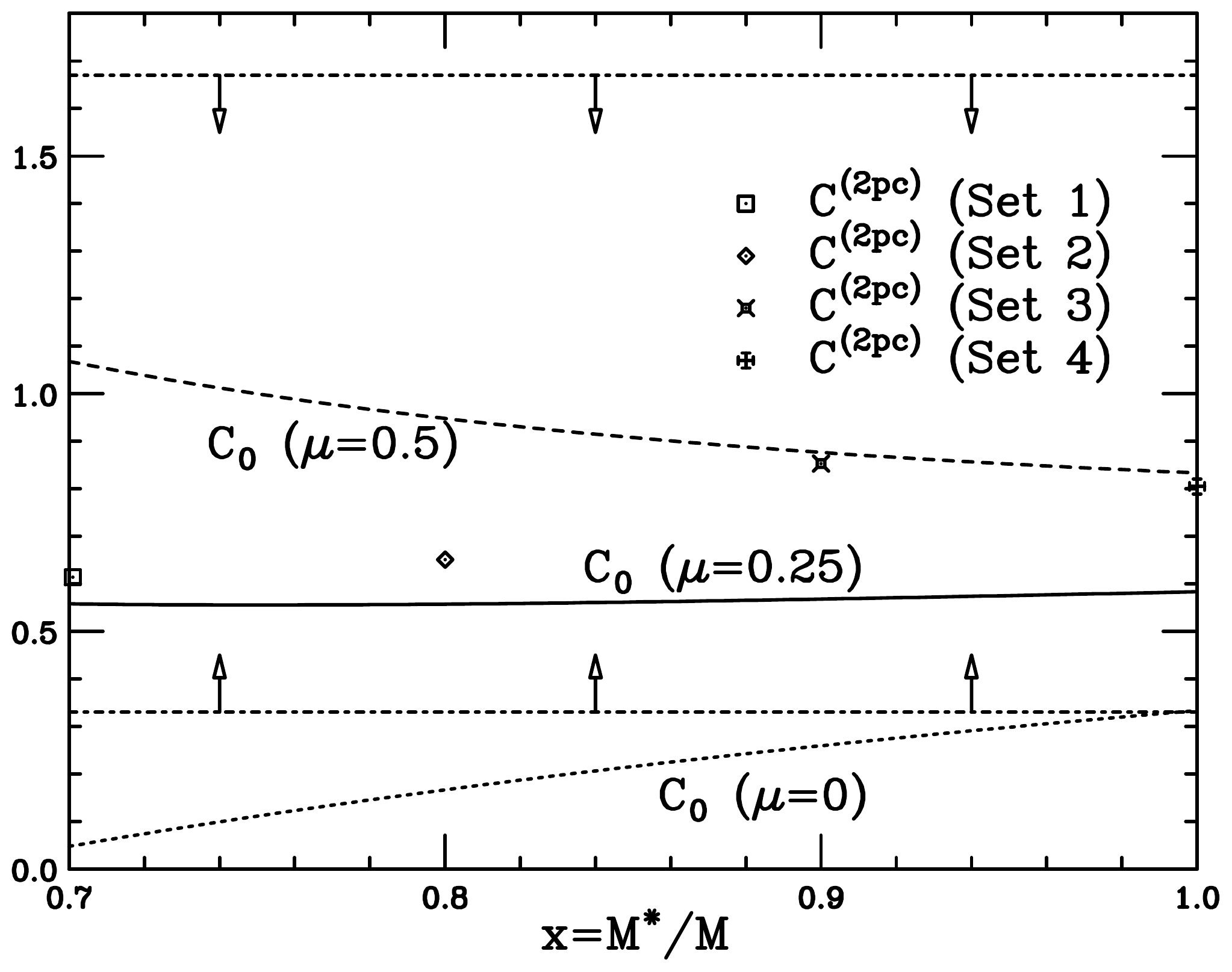}
\caption{The dotted, solid, and dashed lines show $C_0(x)$ defined in Eq.~(\ref{c0})
for three values of the quantity $\mu$, defined in Eq.~(\ref{mu}). The horizontal dash-dotted
lines mark the lower and upper bounds of the inequality in Eq.~(\ref{lim}).
The results for $C^{(\rm 2pc)}$, which refer to the approximation given in Eq.~(\ref{ca}) for  
the effective interactions $1 \sim 4$ discussed in the text, are also indicated
by the symbols at the respective values of $M^*/M$.
(See also Tab.~\ref{tab:1} for more details.)} 
\label{fig:plot}
\end{center}
\end{figure}

In order to discuss our general relation, given in Eq.~(\ref{result}), in the light of the above empirical
information, let us express it at normal nuclear matter density ($p^2/2M = 36$ MeV) in
terms of a quantity $C = C_0 + \Delta C$ in the following way:
\begin{align}
3 a_s - L &= 36 \,\left(C_0 + \Delta C \right)~{\rm MeV} \,. 
\label{ll}
\end{align}
Here $C_0$ and $\Delta C$ are defined as   
\begin{align}
C_0 &=  \left(1 - \frac{2}{3} \frac{M}{M^*}\right) + \mu \left(\frac{M}{M^*}\right)^2 \,,
\label{c0} \\
\Delta C &= - \frac{M}{M^*} \left(H_0' - \frac{1}{3} H_1 \right)\,.
\label{dc} 
\end{align}
The empirical values given in Eq.~(\ref{emp}) imply that $(3 a_s - L)$ is between
$12$ and $60$ MeV. A naive application of these limits to Eq.~(\ref{ll}) gives 
\begin{align}
\frac{1}{3} < \left(C_0 + \Delta C \right) < \frac{5}{3} \,.  
\label{lim}
\end{align}
Let us first consider the possible values of $C_0$ for the range $0 < \mu < 0.5$ and $M^*/M$ given by Eq.~(\ref{mst}). 
Fig.~\ref{fig:plot} shows $C_0$ as a function of $x=M^*/M$ for the case of small $\mu$ ($\mu = 0$), medium $\mu$ ($\mu=0.25$), and large $\mu$ ($\mu=0.5$).
%, togeher with the lower and upper limits of the inequality (\ref{lim})
For small values of $\mu$, $C_0$ 
can take values between $\simeq 0.05$ and $\simeq \frac{1}{3}$, indicating clearly the need
of the three-particle term $\Delta C$ to satisfy Eq.~(\ref{lim}). For intermediate values of $\mu$, $C_0$ is a very slowly varying function of $x$ with values $\lesssim 0.6$, which also would suggest the need of the 
three-particle term if the actual value of $3 a_s - L$ turns out to exceed $\sim 0.6 \times 36 \simeq 22$ MeV. For large values of $\mu$, $C_0$ can take values up to $\sim 1.07$, and the three-particle term is needed only if the actual value of $3 a_s - L$ would turn out to exceed $\sim 1.07 \times 36 \simeq 39$ MeV.

%===============================================================================
%===============================================================================

In the literature~\cite{Goriely:2010bm,Li:2018lpy}, the parameter $\mu$ is often associated with an
``isovector effective mass'' ($M^*_V$), which in turn is related to the 
Landau-Migdal parameter $F_1'$. In App.~\ref{app:A}, we give a detailed discussion on 
this point.\footnote{As shown in App.~\ref{app:A}, the exact relation between $\mu$ and the interaction parameters is far more complicated than Eq.~(\ref{muo}), see Eq.~(\ref{muc}). Nevertheless, for our semi-quantitative discussions, we will assume the validity of Eq.~(\ref{muo}), because it has been reported to be satisfied by various effective interactions~\cite{Goriely:2010bm,Li:2018lpy}.}
Summarizing, it is often assumed that $\mu$ can be expressed as
\begin{align}
\mu \simeq \frac{2}{3} \frac{M^*}{M} \, F_1' \,,     
\label{muo}
\end{align}
which typically leads to values $\mu \simeq 0.2 \sim 0.3$, and $C_0 \simeq 0.5 \sim 0.7$, as will be seen in Tab.~\ref{tab:1}
below.

\begin{figure}[tbp]
\centering
\includegraphics[scale=0.25]{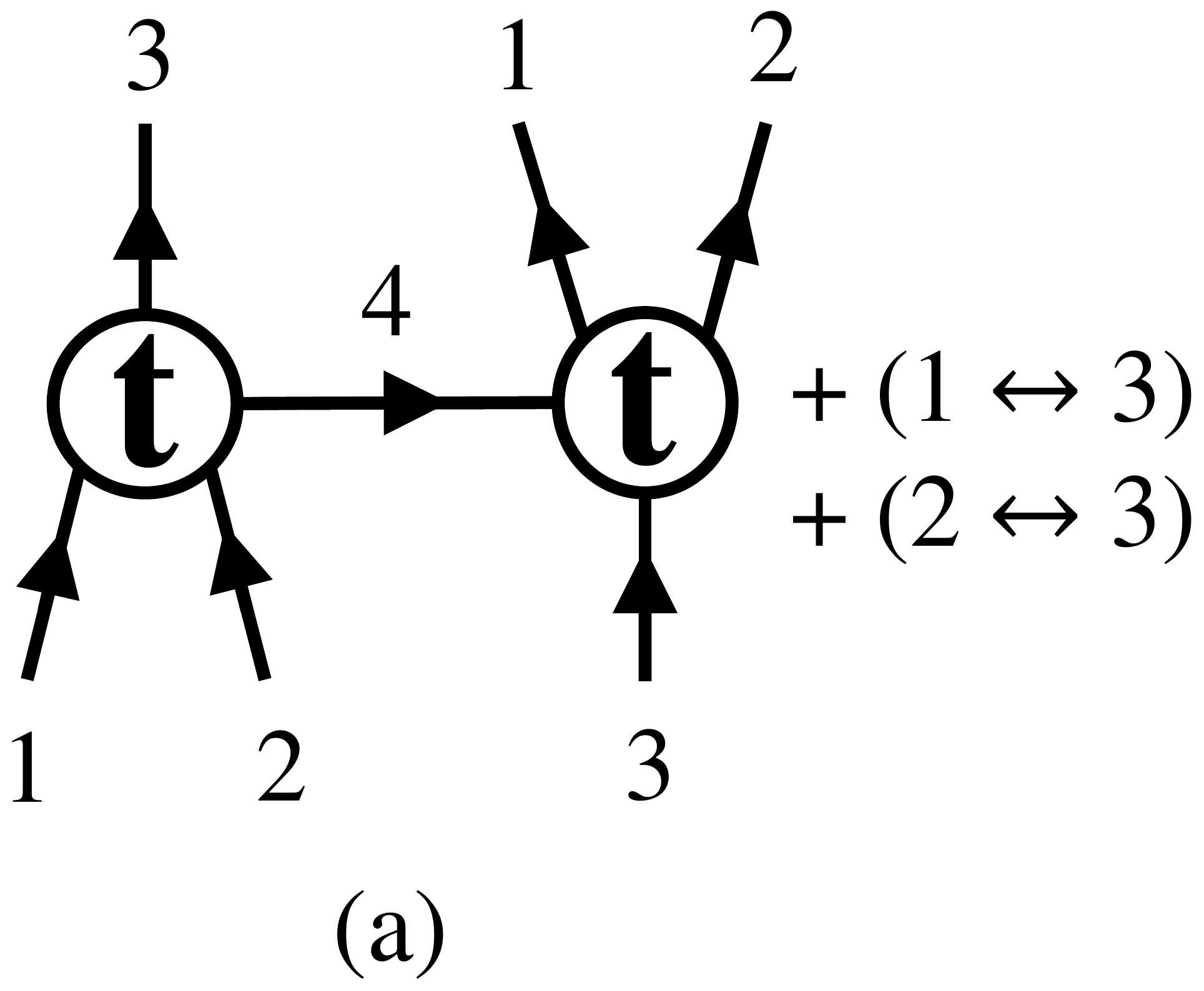}
\hspace{0.5cm} 
\includegraphics[scale=0.2]{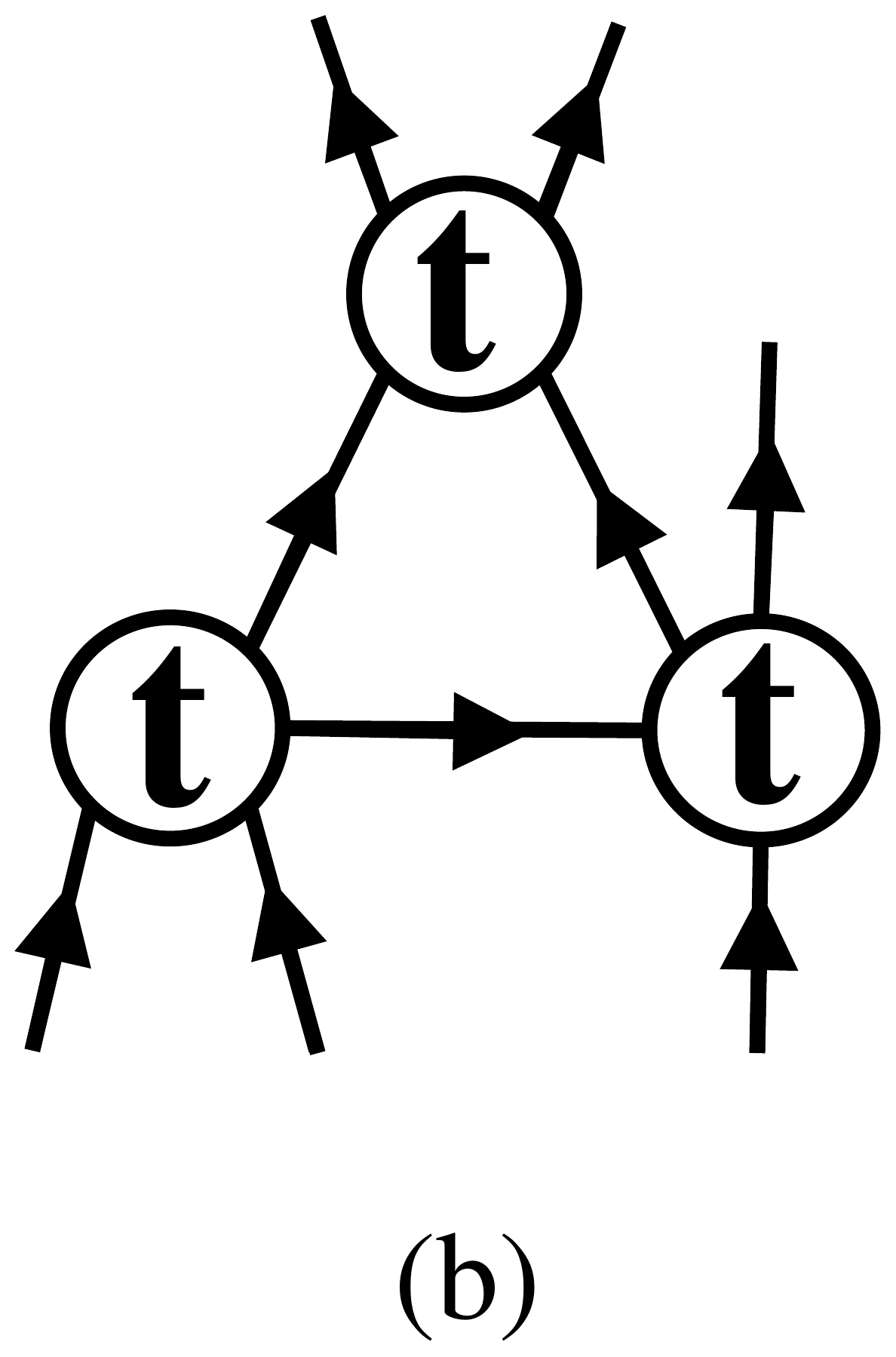}
\caption{(a) The two-particle correlation piece $h^{\rm (2pc)}$ of Eq.~\eqref{expr1}. (b) An example of
the three-particle correlation contribution $h^{\rm (3pc)}$. In each case the solid lines are nucleons, 
and $t$ represents a two-particle scattering matrix. The two diagrams actually represent
the first two terms in the Faddeev series.}
\label{fig:sample}
\end{figure}

Next we wish to address the question of how large the three-particle contribution of Eq.~(\ref{dc}) may be.
For this purpose, we closely follow the semi-quantitative arguments explained in
Ref.~\cite{Bentz:2019lqu}, and split the amplitude $h \equiv h^{(\tau_1 \tau_2 \tau_3)}(\vect{k}_1, \vect{k}_2, \vect{k}_3; \rho)$ in the isospin symmetric limit into a 
two-particle correlation (2pc) piece, a three-particle correlation (3pc) piece, and a residual product piece (prod) according to 
\begin{align}
h = h^{({\rm 2pc})} + h^{({\rm 3pc})} + h^{{(\rm prod)}}.   
\label{split}
\end{align}
The 2pc piece , which is represented by Fig.~\ref{fig:sample}\textcolor{blue}{a}, is
the driving term of the in-medium Faddeev equation, and can be expressed in terms of the two-particle
$t$-matrix by
\footnote{For the derivation, see  Sec.~\ref{sec:III}.}  
\begin{align}
&h^{(\rm 2pc)}(\vect{k}_1, \vect{k}_2, \vect{k}_3; \rho) =
\sum_4 \delta_{\vect{k}_1 + \vect{k}_2, \vect{k}_ 3 + \vect{k}_4} \nonumber \\
&\hs*{12mm}
\times P \, \frac{|\langle 12| \, \hat{t} \, | 34 \rangle_a|^2}
{ \varepsilon_3 + \varepsilon_4 - \varepsilon_1 - \varepsilon_2}  
+ \left( 1 \leftrightarrow 3 \right) +  \left( 2 \leftrightarrow 3 \right). 
\label{expr1}
\end{align}
In this schematic notation, $1 \sim 4$ represent the momenta $\vect{k}_1 \sim \vect{k}_4$
as well as the associated spin and isospin components, though an average over the spin
components of $1, 2, 3$ is assumed implicitly. The sum represents momentum integration
and summation over spin and isospin components of $4$, the $\delta$ symbol represents a momentum
conserving $\delta$-function, $P$ denotes the principal value, 
$\langle 12| \, \hat{t} \, | 34 \rangle_a$ is the antisymmetrized two-particle scattering matrix 
[which is the off-forward generalization of the function
$f$ defined by Eq.~(\ref{vare})], and $\varepsilon_i$ are the quasiparticle energies.
An example for the three-particle correlation contribution $h^{({\rm 3pc})}$, which is
the next term in the Faddeev series, is shown in Fig.~\ref{fig:sample}\textcolor{blue}{b}.
The form of these three-particle cluster terms and associated medium induced processes, as well as 
the origin and the form of the residual product term $h^{(\rm prod)}$, will be derived in  Sec.~\ref{sec:III}.)

%===============================================================================
%===============================================================================

In order to get a rough estimate of $h^{\rm (2pc)}$,
we assume that the two-particle $t$-matrix in Eq.~\eqref{expr1} can be represented by an 
effective contact interaction, i.e., by the in-medium scattering length~\cite{Fetter:1971aa}. 
In this case, the angular averages of Eq.~\eqref{old4} concern only the energy denominator of 
Eq.~\eqref{expr1}, and with the further assumption that the
quasiparticle energies can be approximated as $\varepsilon_i = \frac{\vect{k}_i^2}{2 M^*}$,
where $M^*$ is in the range given by Eq.~(\ref{mst}), the angular integrals can be carried out
analytically, with the very simple results~\cite{Bentz:2019lqu}
\begin{align}
&\int \frac{{\rm d} \Omega_{2}}{4 \pi} \,\int \frac{{\rm d} \Omega_{3}}{4 \pi}
\left(\frac{P}{ \varepsilon_3 + \varepsilon_4
-\varepsilon_1 -\varepsilon_2}  + (1 \leftrightarrow 3) + (2 \leftrightarrow 3) \right) \nonumber \\ 
&= \frac{M^*}{p^2} \, \left( 3 \ln 2 \right) \,, 
\label{int1} 
\end{align} 
\begin{align}
&3 \, \int \frac{{\rm d} \Omega_{2}}{4 \pi} \,\int \frac{{\rm d} \Omega_{3}}{4 \pi} \,
\left(\vect{\hat k}_1 \cdot \vect{\hat k}_2\right)  \nonumber \\
&\hs*{20mm}
\times \left( \frac{P}{ \varepsilon_3 + 
\varepsilon_4 -\varepsilon_1 -\varepsilon_2 } + (1 \leftrightarrow 3)
+ (2 \leftrightarrow 3) \right) \nonumber  \\
&= - \frac{M^*}{p^2} \, \left(1 - \ln 2 \right)  \,. 
\label{int2}
\end{align}
Because these simple expressions indicate that $\ell=1$ contributions are suppressed by large factors
compared to the $\ell=0$ contributions, we can expect that the magnitude of $H_1/3$ in Eq.~(\ref{dc}) is only
a few percent of the magnitude of $H_0'$. For the purpose of our semi-quantitative estimate of the
2pc to the three-particle amplitude, we can therefore assume that 
\begin{align}
C^{\rm (2pc)} &\equiv C_0 + \Delta C^{\rm (2pc)} \simeq C_0 - \frac{M}{M^*} \, H_0'  \,. 
\label{ca}
\end{align}
To be specific, we assume that the matrix elements $_a\langle 34| \,\hat{t}\, |12\rangle$ 
can be replaced by the $\ell=0$ part of an effective interaction of the Landau-Migdal type~\cite{Migdal:1967aa,Speth:1977aa,Migdal:1990vm,Kamerdzhiev:2003rd}:
\begin{align}
&_a\langle 34 | \, \hat{t} \,|12 \rangle = f_0 \, \left(\delta_{31} \cdot \delta_{42}\right) + f_0' \, 
\left(\vect{\tau}_{31} \cdot \vect{\tau}_{42} \right) \nonumber \\ 
&\hs*{12mm}
+ g_0 \, \left(\vect{\sigma}_{31} \cdot \vect{\sigma}_{42} \right)
+ g_0' \, \left(\vect{\sigma}_{31} \cdot \vect{\sigma}_{42} \right) \, 
\left(\vect{\tau}_{31} \cdot \vect{\tau}_{42} \right),
\label{lm}
\end{align}
where the notation indicates that the spin and isospin operators are defined to act in the particle-hole channel.
As usual, the effect of exchange terms is assumed to be included in the interaction parameters. 
Performing then the spin-isospin sum over 4 as well as the spin averages over 1,\,2,\,3 in Eq.~\eqref{expr1}, 
elementary isospin algebra gives the following results for the isoscalar [see Eq.~(\ref{h0})] and isovector [see Eq.~(\ref{h0p})] amplitudes $H^{\rm (2pc)}_0$ and $H'_0{}^{\rm (2pc)}$:
\begin{align}
H_0^{\rm (2pc)} &= \ln 2 \cdot
\frac{1}{4} \left(F_0^2 + 3 F_0'{}^{2} + 3 G_0^2 + 9 G_0'{}^{2} \right) \,,
\label{scal} \\
H_0'{}^{\rm (2pc)} &= \ln 2~\cdot \nonumber \\
&\hs*{-5mm}
\frac{1}{4} \left(\frac{1}{3} F_0^2 + \frac{4}{3} F_0 F_0'  - \frac{1}{3} F_0'{}^{2}   
+ G_0^2 +  4 G_0 G_0'  -  G_0'{}^{2} \right) \,.
\label{vect}
\end{align}
Eq.~(\ref{scal}) agrees with the result of Ref.~\cite{Bentz:2019lqu}, which was obtained directly by using the isospin
average over $1, 2, 3$, and used to estimate the three-particle contributions to the skewness ($J$)
of nuclear matter. It is positive definite, working in the desired direction to explain the empirical value of 
$J$.\footnote{The values of $H^{\rm (2pc)}_0$ for the sets $1 \sim 4$ of Tab.~\ref{tab:1} are $0.528$, $0.762$,
$1.095$, and $4.700$, respectively. The large value for Set 4 is due to an exceptionally large
value of $G_0'$, see Fig.~10 of Ref.~\cite{Holt:2017uuq}.} 
On the other hand, one can expect that the isovector three-particle parameter of Eq.~(\ref{vect}) is negative, mainly because 
of the terms $- \frac{1}{3} F_0'{}^{2}$ and $-G_0'{}^{2}$. 

For illustrative purposes, we show in Tab.~\ref{tab:1}
the results for three sets of the extended Skyrme interaction~\cite{Zhang:2016aa}, and chiral effective field theory~\cite{Holt:2017uuq}. The values in the last line of Tab.~\ref{tab:1} give the results for
$C^{\rm (2pc)}$ in the approximation expressed by Eq.~(\ref{ca}), and these values are also indicated 
by the symbols in Fig.~\ref{fig:plot}.

%===============================================================================
\begin{table}[tbp]
\addtolength{\tabcolsep}{6.8pt}
\addtolength{\extrarowheight}{2.2pt}
\begin{tabular}{|c||c|c|c|c|}
\hline
 Set               & 1  & 2  &  3  &  4  \\  \hline 
$M^*/M$            &  $0.7$  &  $0.8$  &  $0.9$  &  $1.0$  \\  \hline  
$\mu$              &  $0.233$  & $0.229$  &  $0.360$  &   $0.167$  \\  \hline
$C_0$              &  $0.524$  & $0.525$  &  $0.704$  &   $0.500$  \\  \hline 
%$H_0^{\rm (2pc)}$ &  $0.528$  &  $0.762$  &  $1.095$  &  $4.700$ 
$H_0'{}^{\rm (2pc)}$  &  $-0.063$ &  $-0.101$  & $-0.135$  &  $-0.305$  \\  \hline  
$C^{\rm (2pc)}$      & $0.614$  &  $0.651$  &  $0.853$   &  $0.805$  \\   \hline   
\end{tabular}
\caption{Values of various physical quantities entering Eqs.~(\ref{c0}) and (\ref{dc}) at nuclear matter saturation density.
%($\rho_0=0.155$ fm$^-3$) 
Sets 1 $\sim$ 3 correspond to the results for the extended Skyrme interactions~\cite{Zhang:2016aa} 
eMSL07, eMSL08, eMS09, respectively, and Set 4 corresponds to the results of chiral effective field theory~\cite{Holt:2017uuq}. The values for $\mu$, defined by Eq.~(\ref{mu}), given in this Table refer to the approximate expression Eq.~(\ref{muo}), $C_0$ gives the values of Eq.~(\ref{c0}), $H'_0{}^{\rm (2pc)}$ refers to Eq.~(\ref{vect}), and $C^{\rm (2pc)}$ refers to the approximation expressed by Eq.~(\ref{ca}).} 
\label{tab:1}
\end{table}
%===============================================================================

Comparing the values for $C_0$ and $C^{(\rm 2pc)}$ in Tab.~\ref{tab:1}, we see that the two-particle correlation contributions 
are typically $20 \sim 30\%$ of $C_0$, except for Set 4 because of an exceptionally large value of $G_0'$.
Because all values of $C^{\rm (2pc)}$ shown in Tab.~\ref{tab:1} and Fig.~\ref{fig:plot} are within the limits given by Eq.~(\ref{lim}), 
we can conclude that, given the present experimental uncertainties, the symmetry energy and its slope parameter do not
require the presence of an isovector three-particle correlation piece $H_0'{}^{\rm (3pc)}$. This is in contrast
to the case found for the skewness of nuclear matter ($J$)~\cite{Bentz:2019lqu}, which suggests the presence of 
an appreciable isoscalar three-particle correlation piece $H_0^{\rm (3pc)}$.

%==========================================================================    
%=========================================================================
\section{PHYSICAL CONTENT OF THE THREE-PARTICLE AMPLITUDE\label{sec:III}}
As we have seen in the previous Section, and in Ref.~\cite{Bentz:2019lqu}, the three-particle amplitude defined in
Eq.~(\ref{vare}) is directly related to observables quantities. It is therefore desirable to have more understanding
on the physics contained in this quantity, and on methods to calculate it by using certain approximations.
The aim of this Section is, therefore, to extend the well known discussions on the two-particle amplitude 
in the Fermi-liquid theory~\cite{Landau:1959aa,Nozieres:1964aa,Klemt:1976zz,Poggioli:1976zz} to the three-particle amplitude, thereby deriving Eq.~(\ref{expr1}), represented by Fig.~\ref{fig:sample}\textcolor{blue}{a},
and the expressions for the three-particle cluster, shown by Fig.~\ref{fig:sample}\textcolor{blue}{b}, as well as other medium induced three-particle interactions.  

In this Section no special emphasis will be placed on the isospin dependence, therefore, we will for simplicity
discuss only the case of isospin symmetric nuclear matter with Fermi momentum $p$ and associated Fermi energy $\varepsilon$. We therefore simplify
our notations by removing the isospin labels in Eq.~(\ref{vare}) and indicate only the momentum variables. 
To re-introduce the isospin labels is straightforward but the resulting expressions will not be written out 
in the rest of this paper. 
We will also omit the label for the background densities $\{\rho\}$ in this Section, because all density variations
will refer only to those background densities and not to the momentum variables of external particle lines in
a Feynman diagram. 

The differential forms of Eq.~(\ref{vare}) in this simplified notation are then 
${\displaystyle \varepsilon(\vect{k}) = \frac{\delta E}{\delta n_{\vect{k}}}}$, and
\begin{align}
f(\vect{k}_1, \vect{k}_2) = \frac{\delta \varepsilon(\vect{k}_1)}{\delta n_{\vect{k}_2}} \,,
\,\,\,\,\,\,\,\,\,h(\vect{k}_1, \vect{k}_2, \vect{k}_3)  = 
\frac{\delta f(\vect{k}_1,\vect{k}_2)}{\delta n_{\vect{k}_3}} \,.    
\label{start}
\end{align}   
Before discussing the physical content of the three-particle amplitude $h$, we review some well known
facts about the two-particle amplitude $f$. 

\subsection{Basics of Fermi-liquid theory\label{sec:IIIa}}
It is well known~\cite{Nozieres:1964aa} that the quasiparticle energy in Landau's definition is the pole of the single
particle Green's function, specified in Eq.~(\ref{s}) below, near the Fermi surface, i.e.,
$\varepsilon(\vect{k}) = \varepsilon_{0}(\vect{k}) + \Sigma(\varepsilon(\vect{k}), \vect{k})$, where
$\varepsilon_{0}(\vect{k})$ denotes the free (kinetic) energy, and $\Sigma(k) \equiv \Sigma(k_0, \vect{k})$ is the
self energy.\footnote{In this Section and App.~\ref{app:B}, variables like $k$, $k'$, etc.
denote 4-momentum variables, while in Sec.~\ref{sec:II} they denoted the magnitude of the 3-momentum variables.} 
From this pole condition one obtains
\begin{align}
\frac{\delta \varepsilon(\vect{k}_1)}{\delta n_{\vect{k}_2}} = Z_{\vect{k}_1}  \frac{\delta \Sigma(k_1)}{\delta n_{\vect{k}_2}}
|_{k_{10} = \varepsilon(\vect{k}_1)} \,,    
\label{dedn}
\end{align}
where ${\displaystyle Z_{\vect{k}} = \left(1 - \frac{\partial \Sigma(k)}{\partial k_0}|_{k_0 = \varepsilon(\vect{k})} \right)^{-1}} \equiv \left( 1 - \Sigma'(\vect{k})\right)^{-1}$
is the quasiparticle wave function renormalization factor.

It is now very useful to consider the self energy $\Sigma(k)$ as a functional of the {\em exact} propagator
$S(k)$, i.e., to represent $\Sigma(k)$ by skeleton diagrams {\em without} self energy insertions
~\cite{Nozieres:1964aa,Poggioli:1976zz}. Then the following identity holds:
\begin{align}
\frac{\delta \Sigma(k_1)}{\delta n_{\vect{k}_2}} = \int \frac{{\rm d}^4 k}{(2\pi)^4} \, 
\frac{\delta \Sigma(k_1)}{\delta S(k)} \, \frac{\delta S(k)}{\delta n_{\vect{k}_2}} \,.  
\label{dsidn}
\end{align}
The propagator can in turn be expressed in terms of the self energy by
\begin{align}
S(k) = \frac{1}{ k_0 - \varepsilon_0(\vect{k}) - \Sigma(k) - i \eta \left(2 n_{\vect{k}}-1 \right)},
\label{s}
\end{align}
where $\eta = 0^+$, and the last term in the denominator, which is relevant only near the pole
and near the Fermi surface, is equal
to $i \eta$ for unoccupied (particle) states, and $-i\eta$ for occupied (hole) states. From this expression
one obtains the following important identity~\cite{Nozieres:1964aa,Poggioli:1976zz} (see also App.~\ref{app:B}):
\begin{align}
\frac{\delta S(k)}{\delta n_{\vect{k}'}} &= i \left(2\pi \right)^4 \delta^{(3)}\left(\vect{k}-\vect{k}'\right) 
\,\delta\left(k_0 - \varepsilon(\vect{k})\right)\, Z_{\vect{k}}  \nonumber \\
& + S^2(k) \, \frac{\delta \Sigma(k)}{\delta n_{\vect{k}'}} \,.
\label{dsdn}
\end{align}
Here, the first term is obtained when the functional derivative acts on the explicit dependence on the
distribution function in the denominator of Eq.~(\ref{s}), i.e., it expresses the shift of the 
pole from the lower to the
upper $k_0$ plane when a particle with momentum $\vect{k}' = \vect{k}$ is added to the background.
In the second term of Eq.~\eqref{dsdn}, $S^2(k)$ for the pole part simply means $S_p(k)^2 + S_h(k)^2$,
where $S_p$ and $S_h$ are the particle and hole parts of the propagator, i.e., no product
of pole parts $S_p(k) \, S_h(k)$  (which, in a naive sense, is  proportional to $n_{\vect{k}} (1-n_{\vect{k}}) =0$)  
is involved here. 
We now insert Eq.~\eqref{dsdn} into Eq.~\eqref{dsidn}, and define the off-shell quantity $t(k, k')$ by
\begin{align}
\frac{\delta \Sigma(k)}{\delta n_{\vect{k}'}} \equiv Z_{\vect{k}'} \, t(k, k')|_{k_{0}' = \varepsilon(\vect{k}')} \,.   
\label{deft}
\end{align}
As a result, we obtain
\footnote{Here and in the following, the notation $A(k_1, k_2, \dots)|_{k_{0i} = \varepsilon(\vect{k}_i)}$ 
means that all 4-momenta in $A$ should be taken on their energy shells.}  
\begin{align}
f(\vect{k}_1, \vect{k}_2) = Z_{\vect{k}_1} \, Z_{\vect{k}_2} \, 
t(k_1, k_2)|_{k_{0i} = \varepsilon(\vect{k}_i)} \,.     
\label{ff}
\end{align}
Here $t(k_1, k_2)$ is a solution of the integral equation
\begin{align}
t(k_1, k_2) &= K^{(2)}(k_1, k_2)  \nonumber \\
& - i \int \frac{{\rm d}^4 k}{(2\pi)^4} \, K^{(2)}(k_1, k) \, S^2(k) \, t(k, k_2) \,,
\label{bs1}
\end{align}
where the two-particle kernel $K^{(2)}(k_1, k_2)$ is defined by
\begin{align}
K^{(2)}(k_1, k_2) = i \frac{\delta \Sigma(k_1)}{\delta S(k_2)} \,.     
\label{k2}
\end{align}
By the definition of the functional derivative, the kernel $K^{(2)}(k_1, k_2)$ is symmetric under the exchange of
$k_1$ and $k_2$, and by iteration of Eq.~\eqref{bs1} also $t(k_1, k_2)$ is symmetric.

%===============================================================================
\begin{figure}
\begin{center}
\includegraphics[scale=0.28,angle=0]{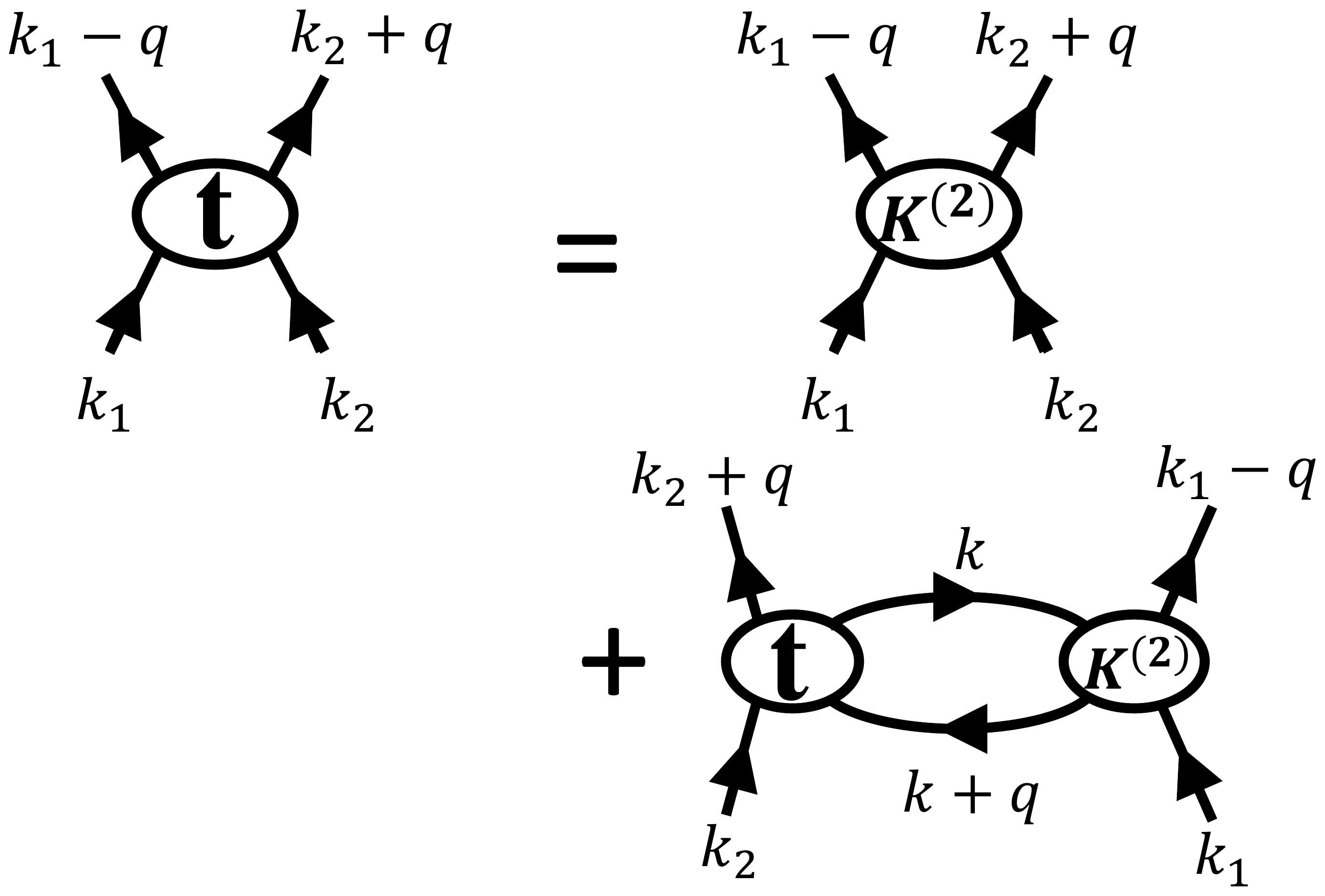}
\caption{Graphical representation of the BS equation in the particle-hole channel for the two-particle t-matrix. Eq.~(\ref{bs1}) corresponds to the forward limit, taking $\vect{q} \rightarrow 0$
before $q_0 \rightarrow 0$.  
} 
\label{fig:bs}
\end{center}
\end{figure}
%===============================================================================

Eq.~(\ref{bs1}), which we represent graphically in Fig.~\ref{fig:bs}, is actually an exact form of the Bethe-Salpeter (BS) equation 
for the two-particle forward scattering amplitude expressed in the particle-hole channel ($t$-channel), because the kernel $K^{(2)}(k_1, k_2)$ is
irreducible in this channel, i.e., it cannot be made disconnected by cutting a pair of lines with the 
same 4-momenta pointing in opposite directions~\cite{Baym:1961zz,Klemt:1976zz}. 
Eq.~(\ref{bs1}) can be derived directly from the definition of the two-particle Green's function 
by using the external field method~\cite{Baym:1961zz}.  
The forward scattering limit is as indicated in the caption to Fig.~\ref{fig:bs}:
If we express the two-particle $t$-matrix generally by $t(k_1', k_2'; k_1, k_2)$, where $k_1$, $k_2$ are the incoming and
$k_1', k_2'$ the outgoing 4-momenta, the forward limit is defined as
\begin{align}
t(k_1, k_2) \equiv  \lim_{q_0 \rightarrow 0} \, \lim_{\vect{q}\rightarrow 0} \, t(k_1+q, k_2-q; k_1, k_2) \,.
\label{fw}
\end{align}
This way of taking the limits, which ensures that $S^2(k)$ in Eq.~(\ref{bs1}) does not involve the product of 
pole parts of particle and hole propagators, defines the quasiparticle interaction in the Fermi-liquid theory
~\cite{Landau:1959aa,Nozieres:1964aa,Negele:1988aa}. We note that, for the case where the propagators are approximated by their pole parts, 
the second term on the r.h.s. of Eq.~\eqref{bs1} contributes only 
if the two-particle $t$-matrix is energy dependent~\cite{Dickhoff:1981wk}.

\subsection{General form of the three-particle amplitude\label{sec:IIIb}}
In order to calculate the three-particle amplitude from Eq.~(\ref{start}), we have to take the functional
derivative of Eq.~\eqref{ff}. The functional derivatives of the $Z$-factors and of the two-particle amplitude w.r.t. 
the energy variables 
%($k_{10}=\epsilon({\vect{k}_1)$ and $k_{20}=\epsilon({\vect{k}_2$) 
give rise to terms which have the form of products of functions 
depending only on two momentum variables. We will call those terms ``product terms'', see 
Eq.~(\ref{hprod}) below for the final form.  They arise
from the energy dependence of the self energy and the two-particle $t$-matrix. For example, by using the
definition of the $Z$-factors given below Eq.~(\ref{dedn}), and Eq.~(\ref{deft}), we have  
\footnote{The term $\frac{1}{2}\left(Z_{\vect{k}}^2 \, \Sigma''(\vect{k}) \right)$ is actually the
second term in the Laurent expansion of the propagator Eq.~\eqref{s} around the pole, i.e.,
$S(k) = \frac{Z_{\vect{k}}}{k_0 - \varepsilon(\vect{k})} + \frac{1}{2} \left(Z_{\vect{k}}^2 \, \Sigma''(\vect{k}) \right)
+ {\cal{O}} \left(k_0 - \varepsilon(\vect{k})\right)$.} 
\begin{align}
\frac{\delta Z_{\vect{k}_1}}{\delta n_{\vect{k}_3}} &= Z_{\vect{k}_1}^{2} \,Z_{\vect{k}_3} 
\frac{\partial t(k_1, k_3)}{\partial k_{10}}|_{k_{i0} = \varepsilon(\vect{k}_i)}
\nonumber \\  
&\hs*{30mm}
+ \left(Z_{\vect{k}_1}^2 \, \Sigma''(\vect{k}_1) \right)\, f(\vect{k}_1, \vect{k}_3) \,,  
\label{dzdn}
\end{align}
and a similar expression for the derivative of $Z_{\vect{k}_2}$,
where we defined 
${\displaystyle \Sigma''(\vect{k}) \equiv \frac{\partial^2 \, \Sigma(k)}{\partial k_0^2}|_{k_0 = \varepsilon(\vect{k})}}$.
Another product term which follows by acting with $\frac{\delta}{\delta n_{\vect{k}_3}}$ on the energy variables
of $t(k_1, k_2)$ in Eq.~\eqref{ff} involves the expression:
\begin{align}
&\frac{\partial t(k_1, k_2)}{\partial k_{10}}|_{k_{i0} = \varepsilon(\vect{k}_i)} \, f(\vect{k}_1, \vect{k}_3) \no \\
&\hs*{38mm}
+ \frac{\partial t(k_1, k_2)}{\partial k_{20}}|_{k_{i0} = \varepsilon(\vect{k}_i)} \, f(\vect{k}_2, \vect{k}_3) \,.
\nonumber
\end{align}
On the Fermi surface ($|\vect{k}_i| = p$, $k_{0i} = \varepsilon$ for $i=1,2,3$), 
the sum of the product terms discussed above can be expressed in the form 
\begin{align}
&\frac{1}{2} \left(\frac{\partial f(\vect{k}_1, \vect{k}_3)}{\partial \varepsilon}  + 
\frac{\partial f(\vect{k}_2, \vect{k}_3)}{\partial \varepsilon} \right) f(\vect{k}_1, \vect{k}_2)  \nonumber \\ 
& + \frac{1}{2} \frac{\partial f(\vect{k}_1, \vect{k}_2)}{\partial \varepsilon}
\left( f(\vect{k}_1, \vect{k}_3) + f(\vect{k}_2, \vect{k}_3) \right)  \nonumber \\
& + \left(Z \, \Sigma'' \right) \, f(\vect{k}_1, \vect{k}_2) \left( f(\vect{k}_1, \vect{k}_3) + f(\vect{k}_2, \vect{k}_3) \right) \,.
\label{part}
\end{align}
Here we used the symmetry of the two-particle amplitude to define, similar to Eq.~(\ref{ddef}) of the previous Section,
\begin{align}
\frac{\partial f(\vect{k}_1, \vect{k}_2)}{\partial \varepsilon} \equiv Z^2 \, 
\left[\left(\frac{\partial}{\partial k_{10}} + \frac{\partial}{\partial k_{20}}\right) t(k_1, k_2) \right]_{k_{i0} = \varepsilon} \,,  
\label{depsnew}
\end{align}
which is the same as the derivative w.r.t. only one energy variable, multiplied by 2.
In Eq.~\eqref{part}, $Z$ and $\Sigma''$ denote the values of $Z_{\vect{k}}$ and $\Sigma''(\vect{k})$ on the Fermi surface.

The product terms Eq.~\eqref{part} are obviously symmetric in $\vect{k}_1$ and $\vect{k}_2$, but do not have a definite
symmetry w.r.t. $\vect{k}_3$. We will see later that additional product terms arise from the functional derivative
${\displaystyle \frac{\delta t(k_1, k_2)}{\delta n_{\vect{k}_3}}}$, evaluated at 
$k_{10} = \varepsilon(\vect{k}_1)$ and
$k_{20} = \varepsilon(\vect{k}_2)$, which make the sum of all product terms totally symmetric in the three momentum variables
$\vect{k}_1, \vect{k}_2, \vect{k}_3$. (See Eq.~(\ref{hprod}) for the final expression.)

\begin{figure}
\begin{center}
\includegraphics[scale=0.40,angle=0]{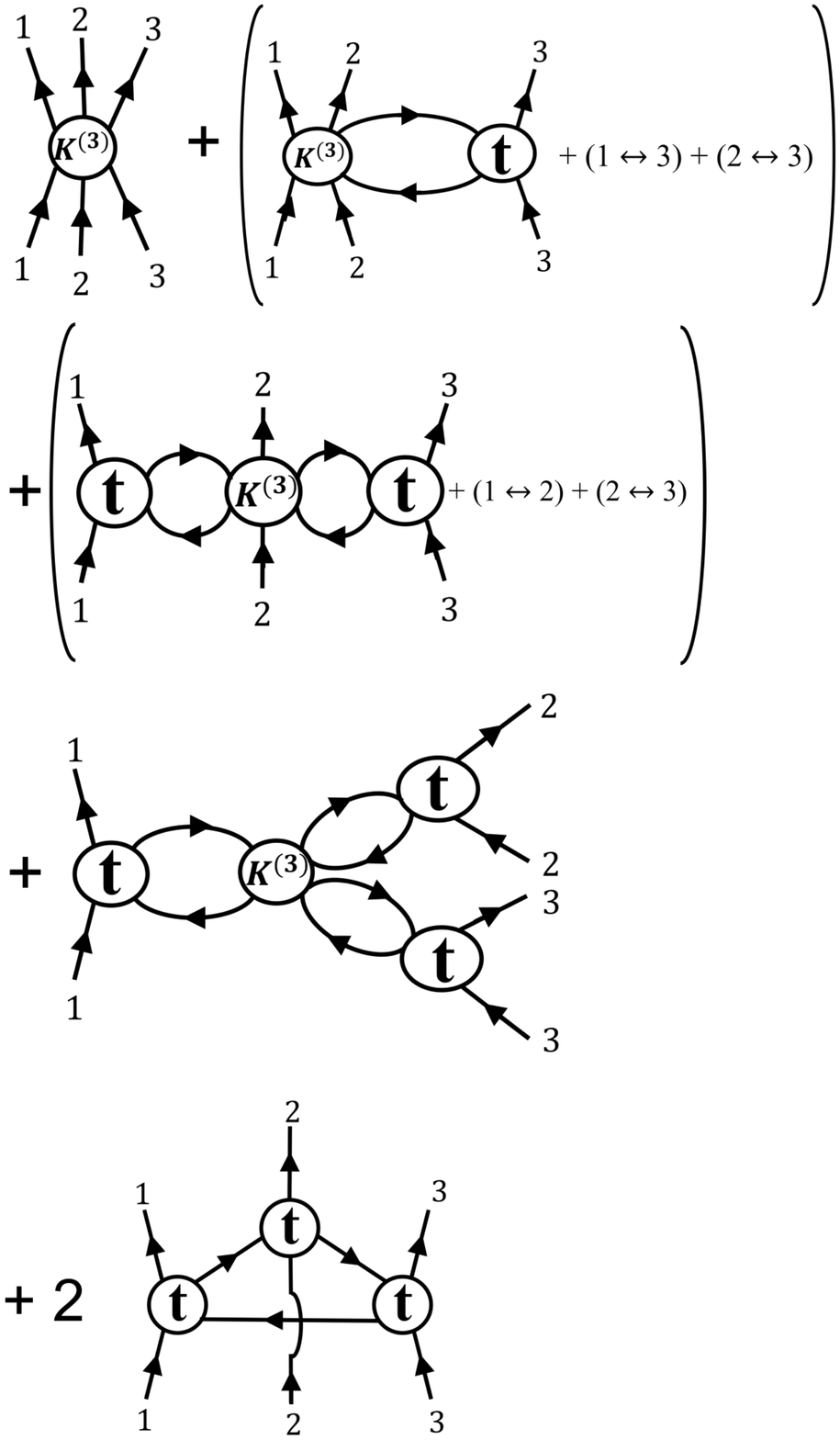}
\caption{Graphical representation of the three-particle amplitude $\tilde{h}(k_1, k_2, k_3)$ of
Eq.~(\ref{fig:ht}). The five diagrams shown here correspond to Eqs.~(\ref{hone})--(\ref{hfive}).} 
\label{fig:ht}
\end{center}
\end{figure}

To calculate ${\displaystyle \frac{\delta t(k_1, k_2)}{\delta n_{\vect{k}_3}}}$, we simply take the functional derivatives of each term
in the BS equation Eq.~\eqref{bs1}. This is in principle the same method as used in Refs.~\cite{Speth:1970aa,Ring:1974egq}.
Consider first the two-particle kernel $K^{(2)}$. If it is expressed by skeleton diagrams (i.e., 
if it is considered as a functional of the exact propagator $S(k)$), we can write down a relation analogous to Eq.~(\ref{dsidn}).
By using Eq.~\eqref{dsdn} and Eq.~\eqref{deft}, this relation can be expressed as
\begin{align}
&\frac{\delta K^{(2)}(k_1, k_2)}{\delta n_{\vect{k}_3}}  = Z_{\vect{k}_3} \, K^{(3)}(k_1, k_2, k_3)|_{k_{30} = \varepsilon(\vect{k}_3)} \nonumber \\
& \hs*{3mm}- i Z_{\vect{k}_3} \, 
\int \frac{{\rm d}^4 k'}{(2\pi)^4} \, K^{(3)}(k_1, k_2, k') \, S^2(k') \, t(k', k_3)|_{k_{30} = \varepsilon(\vect{k}_3)} \,,
\label{dk2dn}
\end{align} 
where we defined the three-particle kernel in analogy with Eq.~\eqref{k2} by~\cite{Speth:1970aa,Ring:1974egq}
\begin{align}
K^{(3)}(k_1, k_2, k_3) = i \frac{\delta^2 \Sigma(k_1)}{\delta S(k_2) \, \delta S(k_3)} \,.     
\label{k3}
\end{align}
By the property of the functional derivative, this is totally symmetric in $k_1, k_2, k_3$. In analogy to $K^{(2)}$,  
it is that part of the forward three-particle scattering amplitude which 
%is irreducible in the particle-hole channel ($t$-channel), i.e., it
cannot be made disconnected by cutting a pair of lines with the same 4-momenta pointing in opposite directions. (This property will become apparent from the final expression, shown graphically
in Fig.~\ref{fig:ht}.) 

Special care has to be taken for the factor $S^2(k)$ in Eq.~(\ref{bs1}), because a naive application of Eq.~\eqref{dsdn}
%to ${\displaystyle \frac{\delta S^2(k)}{\delta n_{\vect{k}_3}}}$ 
leads to an ambiguous on-shell pole. 
In App.~\ref{app:B} we derive the following counterpart of Eq.~\eqref{dsdn}:
\begin{align}
\frac{\delta S^2(k)}{\delta n_{\vect{k}'}} &= i \left(2\pi \right)^4 \delta^{(3)}\left(\vect{k}-\vect{k}'\right) 
\,\delta\left(k_0 - \varepsilon(\vect{k})\right)\, Z_{\vect{k}}  \nonumber \\
& \times \left(\frac{\partial}{\partial k_0} + Z_{\vect{k}} \Sigma''(\vect{k}) \right) + 2 S^3(k) \, \frac{\delta \Sigma(k)}{\delta n_{\vect{k}'}} \,.
\label{ds2dn}
\end{align}
Here $\frac{\partial}{\partial k_0}$ acts on all functions, except $\delta(k_0 -\varepsilon(\vect{k}))$, in a loop integral over $k$, 
and the pole part of $S^3(k)$ simply means $S_p^3(k) + S_h^3(k)$, without products of particle and hole parts. As shown in App.~\ref{app:B}, 
the term which involves the factor $\left( \dots \right)$ in the second line of the above equation, gives rise to additional product terms,
which effectively symmetrize Eq.~\eqref{part} in all three momenta.    

Defining now 
the three-particle off-shell quantity $t^{(3)}(k_1, k_2, k_3)$ by
\begin{align}
\frac{\delta t(k_1, k_2)}{\delta n_{\vect{k}_3}} \equiv Z_{\vect{k}_3} \, t^{(3)}(k_1, k_2, k_3)|_{k_{03} = \varepsilon(\vect{k}_3)} \,, 
\label{deft3}
\end{align}
we can then write down an integral equation for $t^{(3)}$, which follows by taking the functional derivatives of each term in the BS equation Eq.~\eqref{bs1}. 
This equation, which has been used in the form of a three-particle response function in coordinate space in
Refs.~\cite{{Speth:1970aa,Ring:1974egq}}, is given in Eq.~(\ref{bs2}) of 
App.~\ref{app:B}. Because the kernel of that integral equation is the same as the one in the basic BS equation Eq.~\eqref{bs1}, it can be completely resolved, i.e., expressed in terms of the three-particle kernel
$K^{(3)}$, the two-particle t-martix $t$, and the single particle propagator $S$. (For details, see
App.~\ref{app:B}.)
Because all applications require the three-particle amplitude only for the case 
where the particles are on the Fermi surface ($|\vect{k}_i| = p$, $\varepsilon(\vect{k}_i) = \varepsilon$, where $i=1,2,3$), 
we give the expression only for this case.
Separating the product terms from the others, the final general form of the three-particle amplitude on the Fermi surface takes the form
\begin{align}
h(\vect{k}_1, \vect{k}_2, \vect{k}_3) = h^{(\rm prod)}(\vect{k}_1, \vect{k}_2, \vect{k}_3) + 
\tilde{h}(\vect{k}_1, \vect{k}_2, \vect{k}_3) \,,      
\label{hsplit}
\end{align}
where $h^{(\rm prod)}$ is given by
\begin{align}
h^{(\rm prod)}(\vect{k}_1, \vect{k}_2, \vect{k}_3) &= \nonumber \\
& \hs*{-20mm} \frac{1}{2} \, f(\vect{k}_1, \vect{k}_2)  
\left( \left(Z \, \Sigma'' \right) +  \frac{\partial}{\partial \varepsilon} \right) 
\left(f(\vect{k}_1, \vect{k}_3) +  f(\vect{k}_2, \vect{k}_3) \right) \nonumber \\
& + (1 \leftrightarrow 3) +  (2 \leftrightarrow 3)\,, 
\label{hprod}
\end{align}
and $\tilde{h}$ is given by
\begin{align}
\tilde{h}(\vect{k}_1, \vect{k}_2, \vect{k}_3) = Z^3  \, 
\tilde{h}(k_1, k_2, k_3)|_{k_{i0} = \varepsilon} \,,   \label{htilde}
\end{align}
with the off-shell three-particle amplitude
\begin{align}
&\tilde{h}({k}_1, {k}_2, {k}_3) = K^{(3)}(k_1, k_2, k_3)  
\label{hone} \\
& -i \int \frac{{\rm d}^4 k}{(2\pi)^4} \left[ t(k_3, k) \, S^2(k) \,  K^{(3)}(k, k_1, k_2)  \right. \nonumber \\  
& \hs*{5mm} \left. + (1 \leftrightarrow 3) +  (2 \leftrightarrow 3) \right]  
\label{htwo} \\
& - \int \frac{{\rm d}^4 k}{(2\pi)^4}  \int \frac{{\rm d}^4 k'}{(2\pi)^4}\, \left[ t(k_3, k') \, S^2(k') \, 
K^{(3)}(k', k_2, k) \right. \nonumber \\ 
& \hs*{5mm} \left. \times S^2(k) \, t(k, k_1) + (1 \leftrightarrow 2) +  (2 \leftrightarrow 3) \right]  
\label{hthree} \\
& +i \int \frac{{\rm d}^4 k}{(2\pi)^4}  \int \frac{{\rm d}^4 k'}{(2\pi)^4}\,  \int \frac{{\rm d}^4 k''}{(2\pi)^4}
\left[ t(k_3, k'') \, S^2(k'') \right. \nonumber \\
& \hs*{3mm} \left. \times t(k_2,k') \, S^2(k') \, K^{(3)}(k'', k', k) \, S^2(k) \, t(k, k_1) \right] 
\label{hfour} \\
& - 2 i  \int \frac{{\rm d}^4 k}{(2\pi)^4} \left[S^3(k) \, t(k_3, k) \, t(k_2, k) \, t(k_1, k) \right] \,.
\label{hfive}
\end{align}
The 5 terms in Eqs.~\eqref{hone}--\eqref{hfive} are graphically represented in Fig.~\ref{fig:ht}.
%We note that, for case where the propagators are approximated by their pole parts, 
%the second term on the r.h.s. of Eq.~\eqref{bs1} contributes only 
%if the two-particle $t$-matrix is energy dependent. 
Like in Eq.~(\ref{bs1}), if the propagators are approximated by their pole parts, all loop integrals in the above expression 
for $\tilde{h}$ are non-zero only if the two-particle
t-matrix and/or the three-particle kernel depend on the energy variables.  

%===============================================================================
%===============================================================================

To summarize this Subsection, we used the basic definition given in Eq.~\eqref{start} to express the three-particle amplitude $h$ by the 
three-particle kernel of Eq.~(\ref{k3}), the two-particle $t$-matrix, and
the single particle propagator. The results are summarized by Eqs.~(\ref{hsplit})--(\ref{hfive}).

\subsection{Ladder approximation as a building block\label{sec:IIIc}}
In order to specify the two-particle and three-particles kernels of Eqs.~(\ref{k2}) and (\ref{k3}), one needs to model the 
functional dependence of the self energy $\Sigma$ on the single-particle propagator. One model which has been widely used
in the literature since the works of Brueckner, Day, Bethe and others~\cite{Brueckner:1955zzb,Day:1967zza,Bethe:1971xm},
is to express it in terms of the two-particle $t$-matrix calculated in the ladder approximation to the BS equation. 
%where the two Fermion lines 
%forming the ladder point in the same direction (both forward or backward in the time direction). 
The aim of this Subsection
is to use this approximation to derive the well known three-particle processes shown in Fig.~\ref{fig:sample}, as
well as associated medium induced correlations of the same order, that is, of third order in $t_1$, where  
$t_1$ denotes the $t$-matrix in ladder approximation, in the framework of the Fermi-liquid theory. 

\begin{figure}
\begin{center}
\includegraphics[scale=0.3,angle=0]{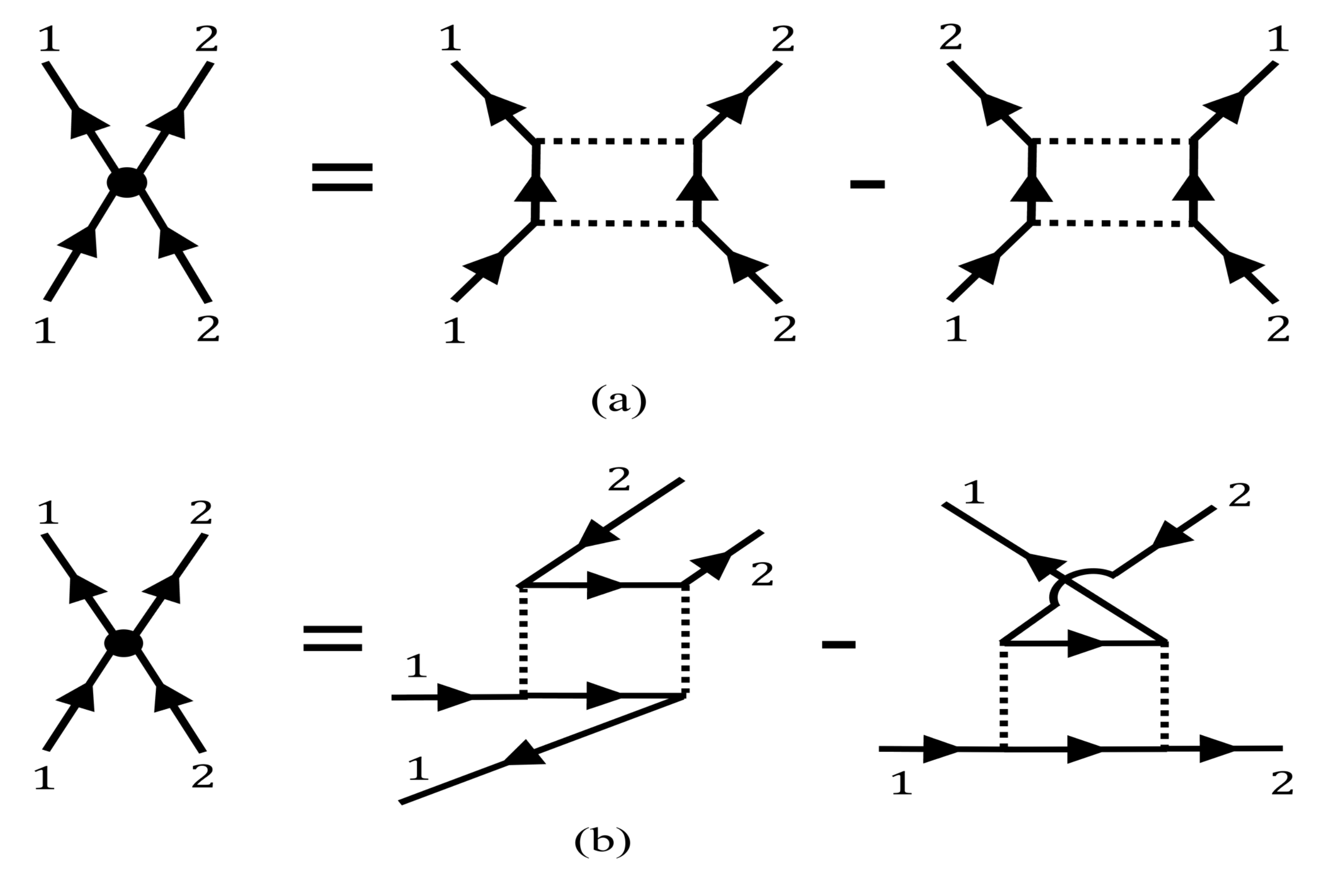}
\caption{(a) Graphical representation of the forward two-particle $t$-matrix in ladder approximation
($t_1$) in the particle-particle channel, where time can be visualized to run from bottom to top.
(b) Graphical representation of $t_1$ in the particle-hole channel, where time can be visualized to run
from left to right. 
The black dot represents $t_1$, and the ladder runs between the two dashed lines. The numbers represent
the 4-momenta of the external particles. The intermediate 
lines in the ladder can be either particles (running forward in time) or holes (running backward
in time).} 
\label{fig:ladder}
\end{center}
\end{figure}

Before proceeding with the formalism, we recall how to visualize the two-particle $t$-matrix in the ladder approximation in the particle-particle channel ($s$-channel, see
Fig.~\ref{fig:ladder}\textcolor{blue}{a}), and in the particle-hole channel ($t$-channel, see Fig.~\ref{fig:ladder}\textcolor{blue}{b}).
Following conventions, the first diagrams in those figures are called the
``direct terms'', while the second are called the ``exchange terms''. We find it convenient to 
include both the
forward and backward propagation in the ladder diagrams, because this avoids the distinction between particles and holes in  subsequent expressions. Nevertheless, except for the minor contributions from backward propagation, our quantity $t_1$ is essentially 
the same as Brueckner's G-matrix~\cite{Brueckner:1955zzb,Bethe:1971xm}.

%===============================================================================
%===============================================================================

The BS equation in the ladder approximation can be expressed for the off-forward case by\footnote{In order to simplify the formulas in this Subsection, 
we denote the external 4-momenta by $1, 1', 2, 2'$ etc., and omit the integral
signs ($\int \frac{{\rm d}^4 k}{(2\pi)^4}$) over internal 4-momenta $\bar{2}, \bar{3}$ etc, which are characterized by a bar over them. 
The $\delta$-functions for
4-momentum conservation in loop integrals are also omitted, as they are evident from the
conservation of total momentum in the $t$-matrix or the bare potential.}
\begin{align}
t_1(1', 2'; 1, 2) &= v(1', 2'; 1, 2) \nonumber \\
& \hs*{-5mm} + \frac{i}{2} \, v(1',2'; \bar{3}, \bar{4}) \, S(\bar{3}) \, S(\bar{4}) \,
t_1(\bar{3}, \bar{4}; 1, 2)\,.    
\label{bs3} 
\end{align}
Here $v$ denotes the static two-particle potential which appears in the underlying Hamiltonian of non-relativistic
field theory~\cite{Fetter:1971aa}. It is assumed to  be antisymmetrized in the incoming (or equivalently the outgoing) particles, i.e., in the operator notation used in Eq.~(\ref{expr1}) 
$v(1', 2'; 1, 2) = \langle 1', 2'| \hat{v}|1,2 \rangle_a$, and similar for $t_1$.
Because we assume static potentials, the BS equation Eq.~\eqref{bs3} can be reduced to a 3-dimensional integral
equation, and $t_1$ actually depends only on the total 4-momentum and two relative 3-momenta, but we will
keep the 4-dimensional notation for clarity.
  
\begin{figure}
\begin{center}
\includegraphics[scale=0.25,angle=0]{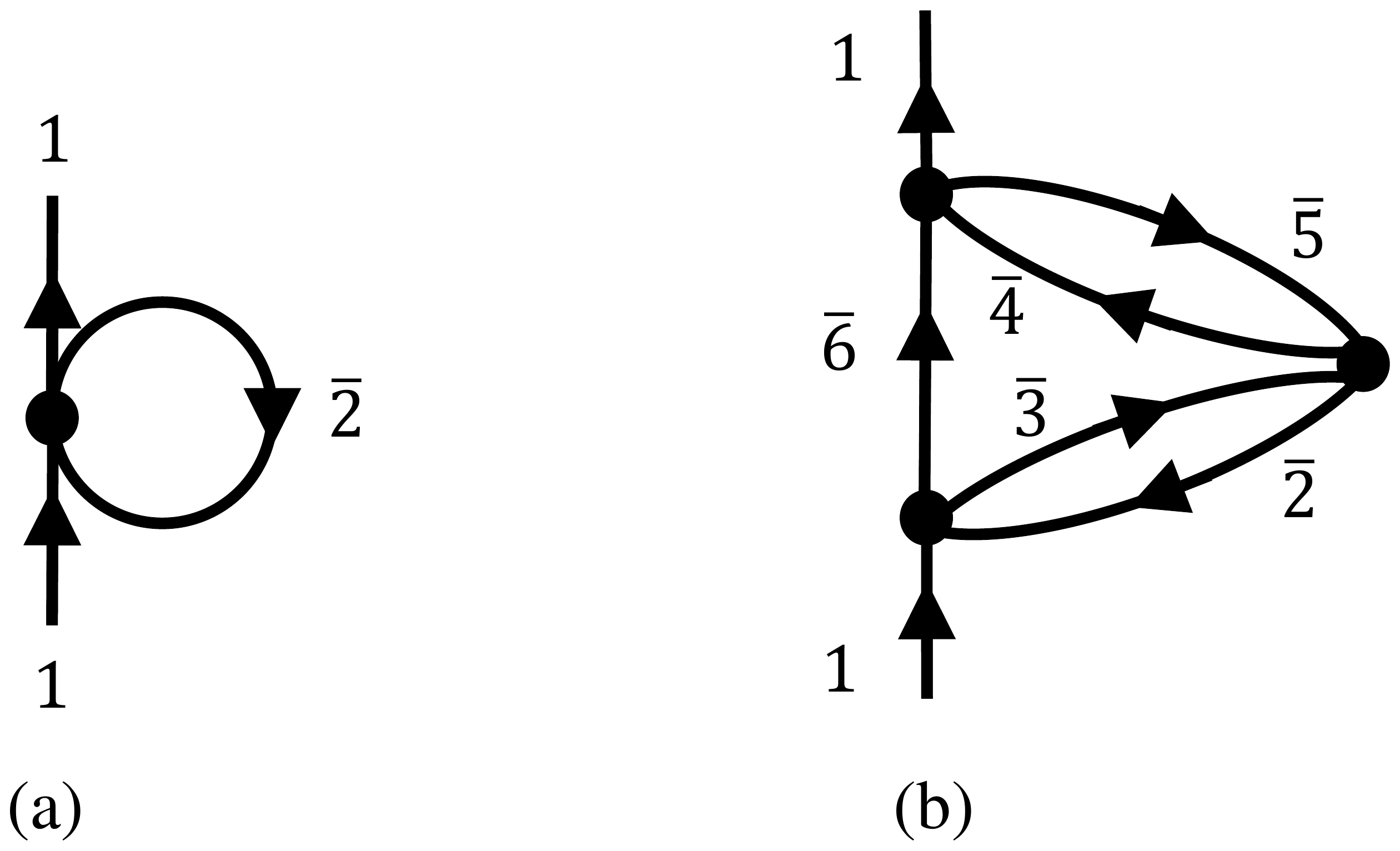}
\caption{(a) Graphical representation of Eq.~(\ref{s1}). (b) Graphical representation of
Eq.~\eqref{s3}. The black dot represents the t-matrix in ladder approximation ($t_1$), and the
numbers represent 4-momenta of the particles. Those with a bar on it refer to integration
variables.} 
\label{fig:se}
\end{center}
\end{figure}

The functional derivative of $t_1$ is also easily obtained from Eq.~(\ref{bs3}) as
\begin{align}
\frac{\delta t_1(1', 2'; 1,2)}{\delta S(3)} = i \, t_1(1', 2'; 3, \bar{4}) \, S(\bar{4}) \, 
t_1(3, \bar{4}; 1, 2)  \,. 
\label{dtds}
\end{align}
By using the Dyson equation for the single particle Green's function and the definition of the two-particle
Green's function, the self energy can be expressed as~\cite{Baym:1961zz,Klemt:1976zz}
\begin{align}
\Sigma(1) &= -i v(1,\bar{2}; 1, \bar{2}) \, S(\bar{2}) \nonumber \\
&+ \frac{1}{2} \, v(1,\bar{2}; \bar{3}, \bar{4}) \, S(\bar{3}) \, S(\bar{4}) \, t(\bar{3}, \bar{4}; 1, \bar{2}) \, S(\bar{2}) \,.   
\label{start1}
\end{align}
The exact two-body $t$-matrix satisfies the BS equation of Eq.~(\ref{bs1}) in the particle-hole channel, 
extended to non-forward kinematics
as indicated already in Fig.~\ref{fig:bs}. In the compact notation used here, it reads    
\begin{align}
t(1', 2'; 1, 2) &= K^{(2)}(1', 2'; 1, 2)  \nonumber \\
& \hs*{-5mm} - i K^{(2)}(1', \bar{3}; 1, \bar{4}) \, S(\bar{3}) \, S(\bar{4}) \, 
t(2', \bar{4}; 2, \bar{3}) \,.   
\label{bs1nf}
\end{align} 
We will use this equation to expand the two-body $t$-matrix in powers of $t_1$ up to the third order:
\begin{align}
t = t_1 + t_2 + t_3 + \dots\,.
\label{tsplit}
\end{align}
To begin, we use the ladder $t$-matrix $t_1$ in Eq.~(\ref{start1}). From the ladder BS equation of Eq.~\eqref{bs3}
we obtain the standard formula for the self energy in Brueckner approximation
\begin{align}
\Sigma_1(1) = -i \, t_1(1,\bar{2}; 1, \bar{2}) \, S(\bar{2}) \,,   
\label{s1}
\end{align}
which is shown as a Hugenholtz diagram in Fig.~\ref{fig:se}\textcolor{blue}{a}. 

%===============================================================================
%===============================================================================

We can use this expression and Eq.~(\ref{dtds}) to calculate the first two terms in the expansion of the
two-body kernel Eq.~\eqref{k2} in powers of $t_1$,
\begin{align}
K^{(2)} = K_1^{(2)} + K_2^{(2)} + K_3^{(2)} + \dots \,,   
\label{k2split}
\end{align}
with the results
\begin{align}
K_1^{(2)}(1,2) &= t_1(1,2; 1,2) \,,
\label{k21}  \\
K_2^{(2)}(1,2) &= i\, t_1(1, \bar{3}; 2, \bar{4}) \, t_1(2, \bar{4}; 1, \bar{3}) \, S(\bar{3}) \, S(\bar{4}) \,,
\label{k22}   
\end{align}
which are symmetric under the interchange $(1 \leftrightarrow 2)$. 
We note that the off-forward generalizations of the kernels given in Eqs.~(\ref{k21}) and (\ref{k22}) are simply
obtained by replacing $1 \rightarrow 1'$ and $2 \rightarrow 2'$ in the final states (the first two
arguments) in each ladder $t$-matrix.  

\begin{figure}
\begin{center}
\includegraphics[scale=0.22,angle=0]{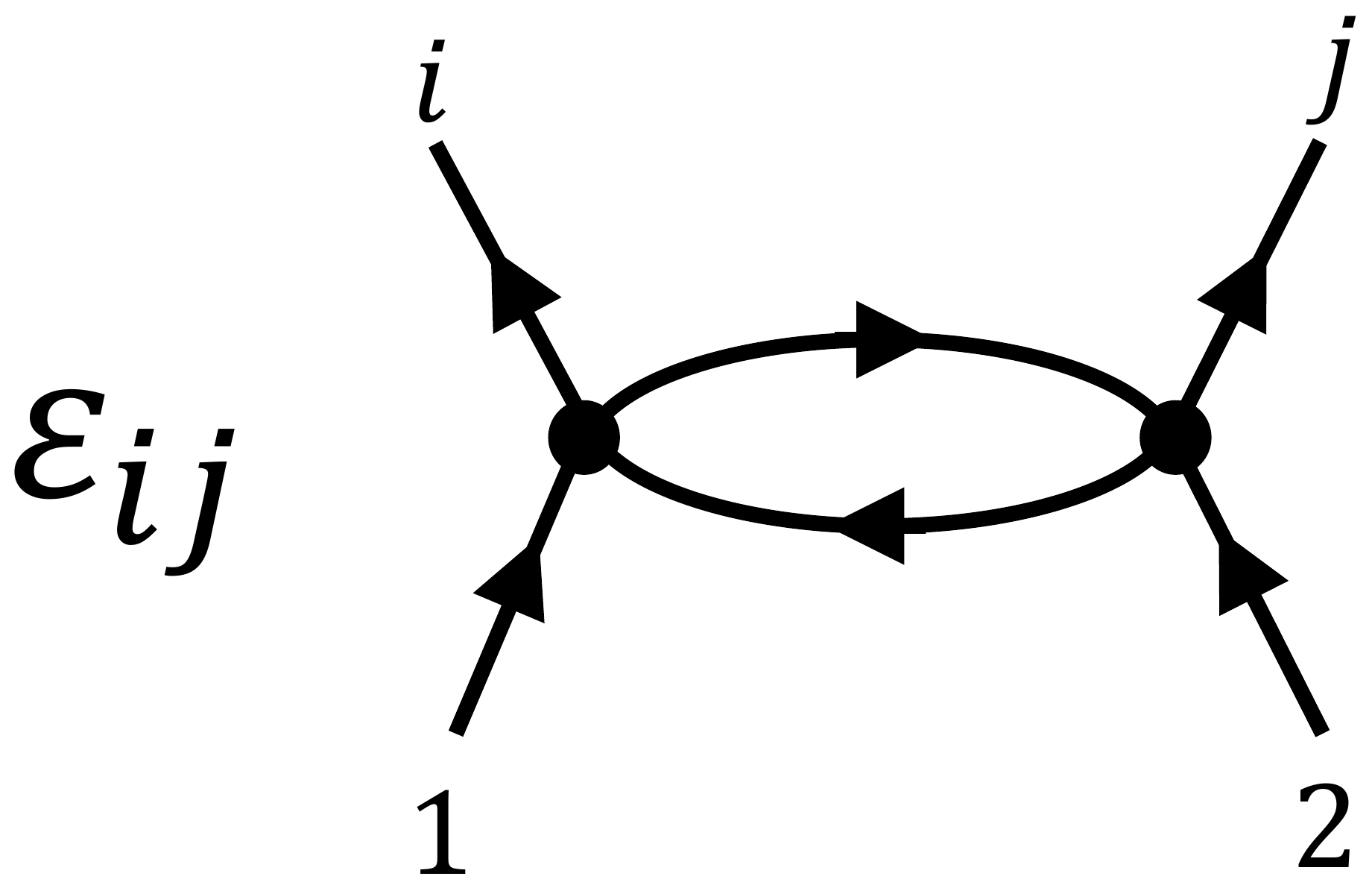}
\caption{Graphical representation of $t_2$ given by Eq.~(\ref{t2nf}) in the forward case.
A sum over $i, j = 1, 2$ is implied, and $\epsilon_{12} = - \epsilon_{21} = 1$.  
For explanation of other symbols, see the caption to Fig.~\ref{fig:se}.}
\label{fig:t2}
\end{center}
\end{figure}

Iterating Eq.~(\ref{bs1nf}) up to the second order in $t_1$, we obtain $t_2$ of Eq.~(\ref{tsplit}):
\begin{align}
t_2(1', 2'; 1, 2) &= K_2^{(2)}(1', 2'; 1, 2) \nonumber \\
& \hs*{0mm} -i K_1^{(2)}(1', \bar{3}; 1, \bar{4})\, S(\bar{3}) \, S(\bar{4}) \, t_1(2', \bar{4}; 2, \bar{3})  \nonumber \\
& \hs*{-15mm} = -i \left[ t_1(1', \bar{3}; 1, \bar{4}) \,  t_1(2', \bar{4}; 2, \bar{3}) 
- (1' \leftrightarrow 2') \right]
S(\bar{3}) \, S(\bar{4}).   
\label{t2nf}
\end{align}
Here, in the last form, the first (direct) term comes from the iteration of Eq.~\eqref{bs1nf}, and
the exchange term comes from $K_2^{(2)}$. We note again that in the forward limit, defined 
by Eq.~(\ref{fw}), there are no contributions from the product $S_p(\bar{3}) S_h(\bar{4})$ in the direct term. 
We represent $t_2$ for the forward case by Fig.~\ref{fig:t2}.

%===============================================================================
%===============================================================================

In passing, we note that by simple contour integration it is easy to confirm that $t_1 + t_2$ in
second order perturbation theory and in the forward limit, where it is equal
to $K_1^{(2)} + K_2^{(2)}$, agrees with the familiar expressions given in Ref.~\cite{Brown:1971zza,Negele:1988aa}.

We now insert $t_2$ of Eq.~\eqref{t2nf} into the second term of Eq.~(\ref{start1}) to get the next term 
in the expansion of the self energy in
powers of $t_1$. For this purpose, we have to eliminate the bare potential $v$ in favor of the ladder $t$-matrix
$t_1$ by using Eq.~\eqref{bs3}. To lowest order, this simply means to replace $v$ by $t_1$ in the
second term of Eq.~\eqref{start1}, and we obtain 
\begin{align}
\Sigma_3(1) &= -i\, t_1(1,\bar{5}; \bar{6}, \bar{4}) \, t_1(\bar{2}, \bar{4}; \bar{3}, \bar{5}) \,
 t_1(\bar{6}, \bar{3}; 1, \bar{2}) \nonumber \\
&\times S(\bar{2}) \, S(\bar{3}) \,  S(\bar{4}) \,  S(\bar{5}) \,  S(\bar{6}) \,,
\label{s3}
\end{align}
which is shown by the Hugenholtz diagram in Fig.~\ref{fig:se}\textcolor{blue}{b}.\footnote{The third order Hugenholtz diagram of Fig.~\ref{fig:se}\textcolor{blue}{b} combines both the rescattering of the 
line $\bar{2}$ in Fig.~\ref{fig:se}\textcolor{blue}{a}, which corresponds to 
Fig.~11.4 of Ref.~\cite{Fetter:1971aa}, and the RPA-type correlations of third order in $t_1$.}
We thus see that there is no second order term of the self energy, i.e.,
\begin{align}
\Sigma(1) = \Sigma_1(1) + \Sigma_3(1) + \Sigma_4 + \dots \,,
\label{sesplit}
\end{align}
and therefore no further contribution to
the second order two-body kernel of Eq.~\eqref{k22}. We now use Eq.~\eqref{s3} to calculate the third order
two-body kernel from Eq.~\eqref{k2} to get
\begin{align}
K_3^{(2)}(1,2) &=  \left[t_1(1, \bar{5}; \bar{6}, \bar{4}) \, t_1(2, \bar{4}; \bar{3}, \bar{5}) \, t_1(\bar{6}, \bar{3}; 1, 2) 
\right.  \nonumber \\
& \hs*{-5mm} \left. +t_1(1, 2; \bar{6}, \bar{4}) \, t_1(\bar{5}, \bar{4}; \bar{3}, 2) \, t_1(\bar{6}, \bar{3}; 1, \bar{5}) \right. \nonumber \\
& \hs*{-5mm} \left. - \left(t_1(1, \bar{5}; \bar{6}, \bar{4}) \, t_1(\bar{3}, \bar{4}; 2, \bar{5}) \, t_1(2, \bar{6}; 1, \bar{3}) 
+ \left(1 \leftrightarrow 2 \right) 
\right)  \right.  \nonumber \\
& \hs*{-5mm} \left. + t_1(1, \bar{5}; 2, \bar{4}) \, t_1(\bar{6}, \bar{4}; \bar{3}, \bar{5}) \, t_1(2, \bar{3}; 1, \bar{6}) \right]
\nonumber \\
& \hs*{-5mm} \times S(\bar{3}) \, S(\bar{4}) \, S(\bar{5}) \, S(\bar{6}) \,,    
\label{k23}
\end{align}
which is also symmetric under the interchange $(1 \leftrightarrow 2)$.

\begin{figure}
\begin{center}
\includegraphics[scale=0.3,angle=0]{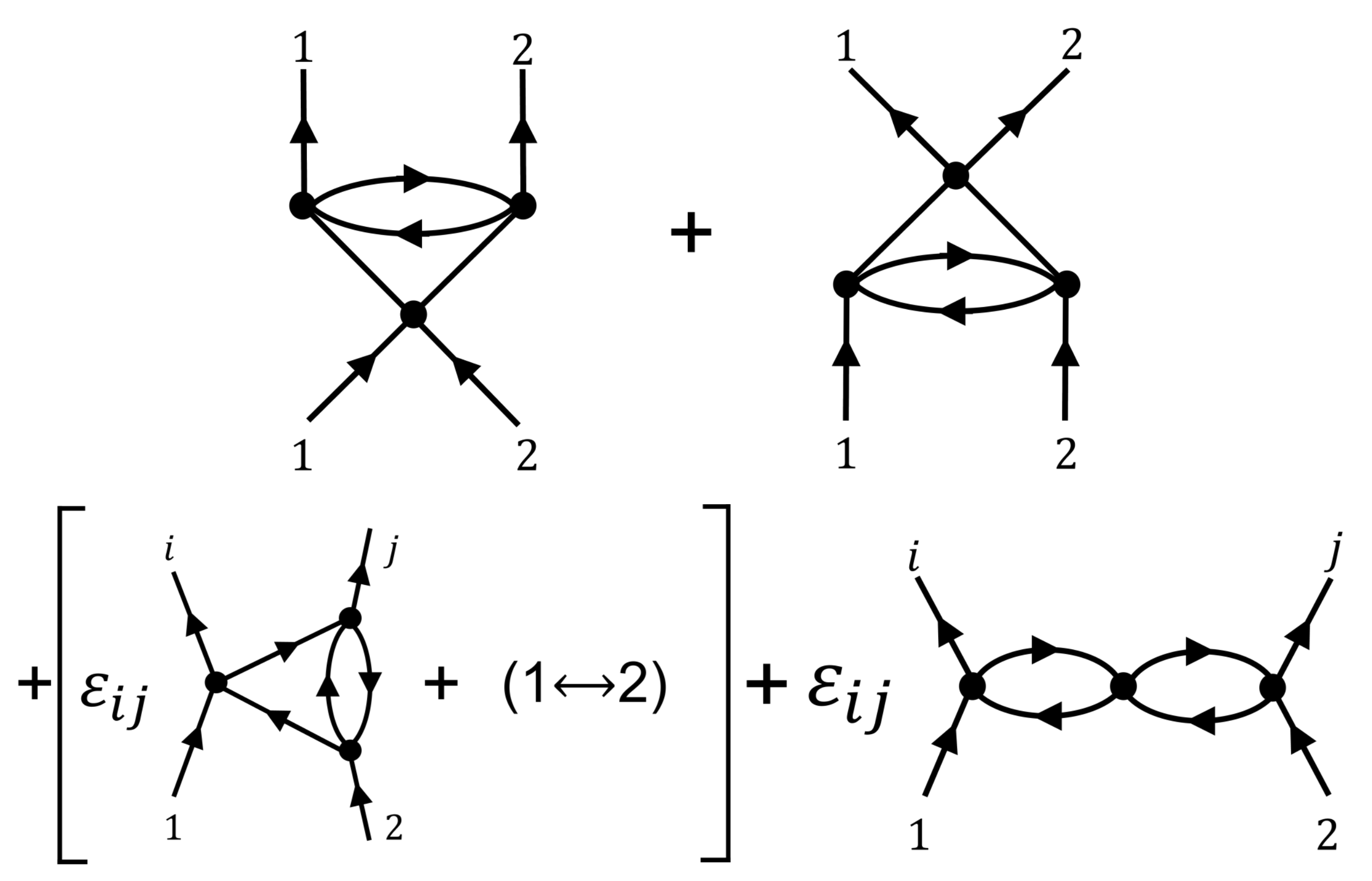}
\caption{Graphical representation of $t_3$ of Eq.~(\ref{t3nf}) in the forward case.
A sum over $i, j = 1, 2$ is implied, and $\epsilon_{12} = - \epsilon_{21} = 1$.
For explanation of other symbols, see the caption to Fig.~\ref{fig:se}.
The three direct terms from $\epsilon_{12} = +1$ correspond to Eqs.~(\ref{second}) and (\ref{third}), 
and the three exchange terms from $\epsilon_{21} = -1$ correspond to the terms in the third and
fourth lines in the kernel of Eq.~(\ref{k23}).}
\label{fig:t3}
\end{center}
\end{figure}

By iteration of (\ref{bs1nf}) we then get for the third order term in Eq.~(\ref{tsplit}):  
\begin{align}
t_3(1', 2'; 1, 2) &= K_3^{(2)}(1', 2'; 1, 2)  \label{first} \\
&\hs*{-10mm}  -i K_2^{(2)}(1', \bar{3}; 1, \bar{4}) \, S(\bar{3}) \, S(\bar{4}) \, t_1(2', \bar{4}; 2, \bar{3}) 
\label{second} \\
&\hs*{-10mm}  -i K_1^{(2)}(1', \bar{3}; 1, \bar{4}) \, S(\bar{3}) \, S(\bar{4}) \, t_2(2', \bar{4}; 2, \bar{3}) \,.
\label{third}
\end{align}
Here the off-forward forms of $K_2^{(2)}$ and $K_3^{(2)}$ are obtained from Eqs.~(\ref{k22}) and (\ref{k23}) 
by replacing $1 \rightarrow 1'$, $2 \rightarrow 2'$ in the final states (first two arguments) of each $t$-matrix,
and $t_2 = K_2^{(2)} -i K_1^{(2)} \,S \, S \, t_1$ is given by Eq.~(\ref{t2nf}).
The explicit form of $t_3$ is given in App.~\ref{app:B}, and the corresponding Feynman diagrams
for the forward case are shown in Fig.~\ref{fig:t3}.
From the above expressions, one can see that the terms Eq.~(\ref{second}) and Eq.~(\ref{third}) are direct terms, with the corresponding
exchange terms given in the third and fourth lines of Eq.~(\ref{k23}) for the off-forward case.  

%===============================================================================
%===============================================================================

In fourth order there are two types of contributions to the self energy Eq.~(\ref{start1}). One is obtained by
inserting $t_3$ into the second term of Eq.~(\ref{start1}) and replacing the bare potential $v$ by $t_1$.
The other, which is the ``rest'' of Eq.~(\ref{s3}), is obtained by using $t_2$ for the $t$-matrix, and the second
order term $ - \frac{i}{2} t_1 S S t_1$ from the elimination of the bare potential $v$ in favor of $t_1$:
\begin{align}
\Sigma_4(1) &= \frac{1}{2}\, t_1(1, \bar{2}; \bar{3}, \bar{4}) \, t_3(\bar{3}, \bar{4}; 1, \bar{2})
\, S(\bar{2}) \, S(\bar{3}) \, S(\bar{4}) \nonumber \\
& - \frac{i}{4} \, t_1(1, \bar{2}; \bar{5}, \bar{6}) \, t_1(\bar{5},\bar{6}; \bar{3}, \bar{4})
t_2(\bar{3},\bar{4} ; 1, \bar{2}) \nonumber \\
& \times S(\bar{2}) \, S(\bar{3}) \, S(\bar{4}) \,  S(\bar{5}) \, S(\bar{6}) \,. 
\label{se4}
\end{align}
The resulting expression is given in App.~\ref{app:B}. We note that the
``counter term'' (second term of Eq.~(\ref{se4})) is necessary to cancel the contribution from the
second diagram of Fig.~\ref{fig:t3} to the first term of Eq.~(\ref{se4}), so as
to avoid double counting of ladder-type contributions. 
%The resulting expression for $\Sigma_4$ is given in App.~\ref{app:B}. 

From $\Sigma_4$ one could now calculate the fourth order two-body kernel by using Eq.~(\ref{k2}),
the fourth order $t$-matrix by using Eq.~(\ref{bs1nf}), the fifth order self energy from Eq.~(\ref{start1}), 
and so on.
Because the main purpose of this Section is to derive the three-particle amplitudes up to third order in $t_1$, we will
not go beyond the third order in the following discussions. Nevertheless, one should keep in mind that 
the formalism can in principle be extended to higher orders, although an algorithm for practical
calculations needs still to be developed.

We now turn to the discussion of the three-particle amplitudes.
From Eq.~(\ref{dtds}) we see that the three-particle kernel $K^{(3)}$ of Eq.~(\ref{k3}) receives terms of order $t_1^2$ from 
$K_1^{(2)}$ and $K_2^{(2)}$, and terms of order $t_1^3$ from $K_2^{(2)}$ and $K_3^{(2)}$.
Thus, up to third order in $t_1$ we can write for the three-particle kernel $K^{(3)}$ and the
corresponding amplitude $\tilde{h}$ of Eqs.~(\ref{hone})--(\ref{hfive}):  
\begin{align}
K^{(3)} &= K_2^{(3)} + K_3^{(3)}, &
\tilde{h} &= \tilde{h}_2 + \tilde{h}_3.  \nonumber
\end{align} 
By using Eq.~(\ref{dtds}), the terms of order $t_1^2$ arising from the functional derivatives of $K_1^{(2)}$ and $K_2^{(2)}$ 
are given by
\begin{align}
\tilde{h}_2(1,2,3) &= K_2^{(3)}(1,2,3)  \nonumber \\
& \hs*{-10mm} = - |t_1(1,2; 3, \bar{4})|^2 \, S(\bar{4}) + \left(1 \leftrightarrow 3\right) +  \left(2 \leftrightarrow 3\right) \,.
\label{k32}
\end{align}
Because of momentum conservation, the 4-momentum corresponding to $\bar{4}$ in the first term of 
this expression has the form 
$k_4 = \left(\varepsilon_1 +\varepsilon_2 - \varepsilon_3 \,, \vect{k}_4 \right)$ with 
$\vect{k}_4 = \vect{k}_1 +  \vect{k}_2 - \vect{k}_3$, and $\varepsilon_i \equiv \varepsilon{(\vect{k}_i)}$. 
Because Eq.~(\ref{k32}) refers to the second order in $t_1$, we can approximate the Feynman propagator $S(\bar{4})$ 
by its pole part with the $Z$-factor replaced by unity (see Eq.~(\ref{s})): 
\begin{align}
& S(\bar{4}) = \frac{1}{ \varepsilon_1 +\varepsilon_2 - \varepsilon_3 - \varepsilon_4 + i \delta \left(2 n_{\vect{k}_4} - 1 \right)}
\nonumber \\
& = \frac{P}{\varepsilon_1 +\varepsilon_2 - \varepsilon_3 - \varepsilon_4} - 
i \pi \delta \left( \varepsilon_1 +\varepsilon_2 - \varepsilon_3 - \varepsilon_4 \right) \left(2 n_{\vect{k}_4} - 1 \right), 
\label{ss4}
\end{align}
where $P$ denotes the principal value. Because we take all three particles on the Fermi surface ($\varepsilon_i = \varepsilon$ for
$i = 1,2,3$), the delta function term implies that also $\varepsilon_4 = \varepsilon$, i.e., $|\vect{k}_4| = p$. 
In order to avoid an unphysical imaginary part of the three-particle amplitude Eq.~(\ref{k32}), one has to define
the step function $n_{\vect{k}} = \theta(p - |\vec{k}|)$ so that for $|\vect{k}| = p$ one has $n_{\vect{k}} = \frac{1}{2}$, which
is also suggested by the zero temperature limit of the Fermi distribution function, going to the Fermi surface before taking the
limit $T \rightarrow 0$. Then the second term in Eq.~(\ref{ss4}) vanishes, and 
$\tilde{h}_2$ of Eq.(\ref{k32}) becomes Eq.~(\ref{expr1}) of the previous section in the ladder approximation
($\hat{t} = \hat{t}_1$). 

The three-particle kernel of order $t_1^3$ is obtained by applying 
$\delta/{\delta S(3)}$ to: (i) $K_2^{(2)}$ by using the relation Eq.~(\ref{dtds}), which gives 2 terms, and (ii)
$K_3^{(2)}$, which gives $4 \times 5 = 20$ terms. These terms in $K_3^{(3)}$ can be divided
into 3 groups A, B, and C. We call class A the traditional Faddeev-type terms represented by 
Fig.~\ref{fig:sample}\textcolor{blue}{b}, 
class B the group of associated medium induced interaction terms to be discussed below, and class
C those diagrams where one of the lowest order vertices $t_1$ in Eq.(\ref{k32}) or Fig.~\ref{fig:sample}\textcolor{blue}{a} is 
replaced by the second order vertex $t_2$ of Eq.~(\ref{t2nf}). 

\begin{figure}
\begin{center}
\includegraphics[scale=0.2,angle=0]{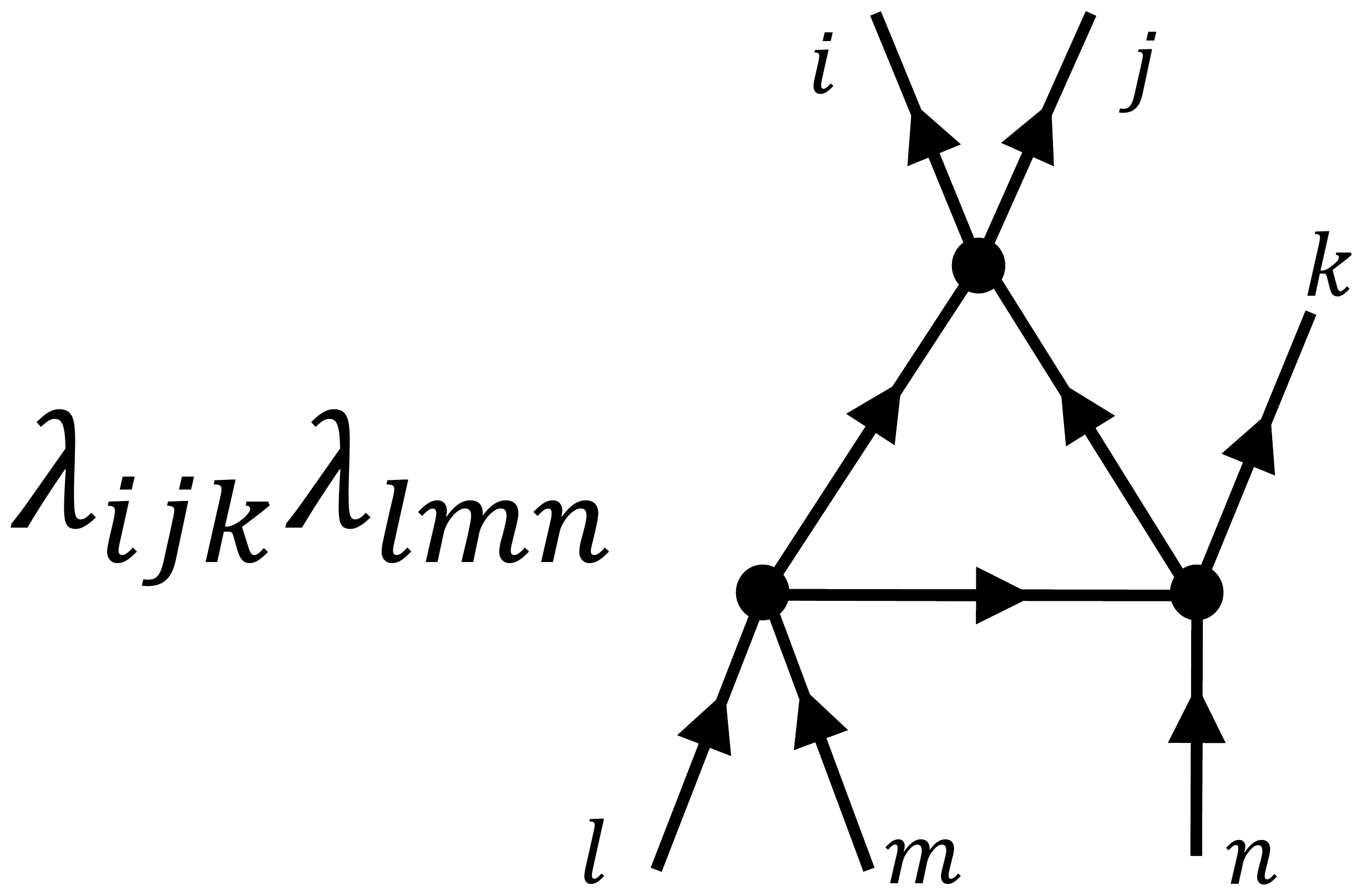}
\caption{Graphical representation of class A three-particle processes, Eq.~(\ref{aa}). 
$\lambda_{i j k}$ is unity for even permutations of
$(1 2 3)$ and zero otherwise, and an independent sum over $(i j k)$ and $(l m n)$ is implied. 
For explanation of other symbols, see the caption to Fig.~\ref{fig:se}.
} 
\label{fig:ta}
\end{center}
\end{figure}

There are 6 terms of type A which arise from $K_3^{(2)}$.
Going back to the self energy $\Sigma_3(1)$ shown in Fig.~\ref{fig:se}\textcolor{blue}{b}, these six terms arise when    
$\delta^2/{\delta S(2) \delta S(3)}$ hits (i) the pair $(\bar{2}, \bar{4})$, (ii) the pair $(\bar{2}, \bar{5})$, and
(iii) the pair $(\bar{3}, \bar{5})$. They are shown graphically by the diagrams of Fig.~\ref{fig:ta}
with $n \neq k$. There are, however, two more contributions to $\tilde{h}$ in third
order, i.e., the terms Eq.~(\ref{htwo}) and Eq.~(\ref{hfive}). 
In our condensed notation, Eq.~(\ref{htwo}) reads to third order
\begin{align}
&- i \, K_2^{(3)}(1,2,\bar{5}) \, S^2(\bar{5}) \, t_1(3, \bar{5} ; 3, \bar{5}) + \left(1 \leftrightarrow 3\right) +
 \left(2 \leftrightarrow 3\right)  \nonumber \\
& = i \left( |t_1(1,2; \bar{4}, \bar{5})|^2 +  |t_1(1,\bar{4}; 2, \bar{5})|^2 + |t_1(1, \bar{5}; 2, \bar{4})|^2
\right) \nonumber \\
& \times t_1(3, \bar{5}; 3, \bar{5}) \, S (\bar{4}) \, S^2(\bar{5}) + 
\left(1 \leftrightarrow 3\right) + \left(2 \leftrightarrow 3\right) \,, 
\label{other}
\end{align}
and Eq.~(\ref{hfive}) reads
\begin{align}
-2 i \, t_1(1,\overline{2}; 1, \overline{2}) \, t_1(2,\overline{2}; 2, \overline{2}) \, 
t_1(3,\overline{2}; 3, \overline{2}) \, S^3(\overline{2}) \,.
\label{other1}
\end{align}
In Eq.~(\ref{other}) we used the form of $K_2^{(3)}$ given in Eq.~(\ref{k32}). 
The first term in Eq.~(\ref{other}), together with the indicated particle exchanges, is also of the Faddeev-type (class A), and is represented by the diagram 
in Fig.~\ref{fig:ta} with $n=k$.  The remaining two terms in 
Eq.~(\ref{other}), as well as Eq.~(\ref{other1}), are of type B and will be discussed later.\footnote{We wish to emphasize again that the pole part of $S^3$ in Eq.~(\ref{other1}) means $S_p^3 + S_h^3$, without any products
$S_p S_h$, and therefore, in the pole approximation for the single particle propagators, Eq.~(\ref{other1}) is non-zero 
only because of the energy dependence of $t_1$.}

The sum of class A contributions to the three-particle amplitude $\tilde{h}$ in third order of the ladder
$t$-matrix can then be compactly expressed by
\begin{align}
\tilde{h}_3^{({\rm A})}(1,2,3) &= i \, \lambda_{i j k} \, \lambda_{l m n} \,
t_1(i, j; \bar{4}, \bar{5}) \,t_1(\bar{4}, \bar{6}; l, m) \nonumber \\
& \times t_1(\bar{5}, k; \bar{6}, n)
S(\bar{4}) \, S(\bar{5}) \, S(\bar{6}) \,,
\label{aa}
\end{align}
shown in Fig.~\ref{fig:ta}. Here we defined the symbol $\lambda_{i j k}$ to be unity for even permutations of
$(1 2 3)$ and zero otherwise, and an independent sum over $(i j k)$ and $(l m n)$ is implied. 
Fig.~\ref{fig:ta} corresponds to the Faddeev decomposition of the three-particle amplitude into a sum over 
all processes where the particle with 4-momentum $n \, (k) = 1, 2, 3$ is the initial (final) spectator.
We note that the amplitude Eq.~(\ref{aa}) is totally symmetric in $(123)$, which corresponds to particle interchanges both
in the initial and final states, but antisymmetric with respect to interchanges of two 4-momenta in the final state,
as required by the Pauli principle.
  
%===============================================================================
%===============================================================================

\begin{figure}
\begin{center}
\includegraphics[scale=0.2,angle=0]{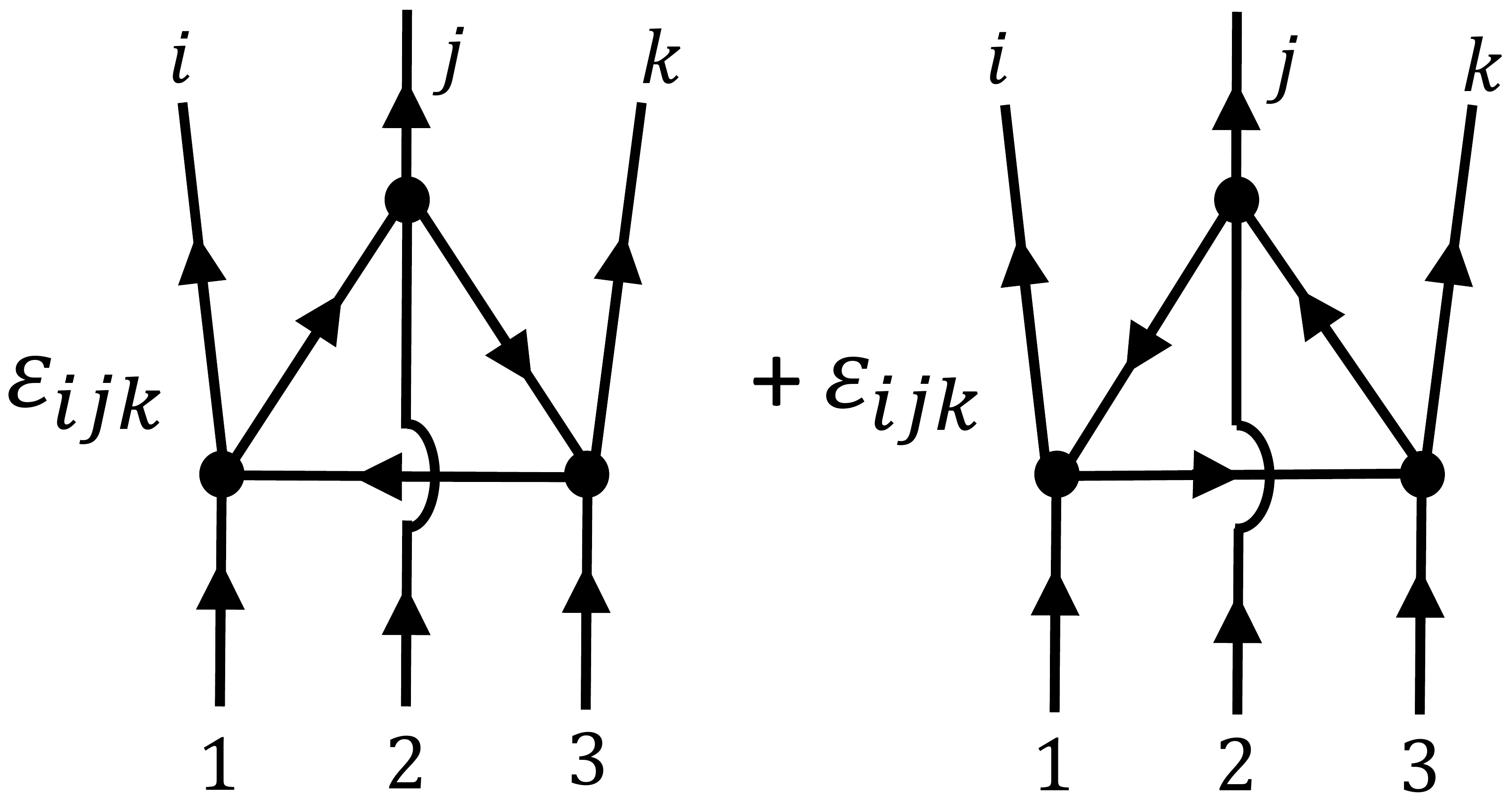}
\caption{Graphical representation of class B three-particle processes. 
$\epsilon_{ijk}$ is the usual antisymmetric tensor, and a sum over $i, j, k = 1, 2, 3$ is implied.
For explanation of other symbols, see the caption to Fig.~\ref{fig:se}.} 
\label{fig:tb}
\end{center}
\end{figure}

The three-particle processes of class B are: (i) 2 terms arising from  
$\delta K_2^{(2)}/{\delta S(3)}$ as mentioned already above; (ii) 2 terms arising from $\delta K_3^{(2)}/{\delta S(3)}$,
which have their origin in cutting the lines $(\bar{3}, \bar{4})$ in the self energy $\Sigma_3$ of
Fig.~\ref{fig:se}\textcolor{blue}{b}, (iii) the second and third terms of Eq.~(\ref{other}), including the indicated particle
exchanges, and (iv) the term Eq.~(\ref{other1}). The latter has a factor of $2$ because the two orientations of the
loop in the last diagram of Fig.~\ref{fig:ht} give the same result, and we therefore consider Eq.~(\ref{other1}) to consist
of 2 identical terms.    
The sum of these 12 three-particle amplitudes of class B in third order of $t_1$ can then be compactly expressed as
\begin{align}
&\tilde{h}_3^{(\rm B)}(1,2,3) = - i \, \epsilon_{ijk} \Big[ t_1(i, \bar{5}; 1, \bar{4}) \,  t_1(j, \bar{6}; 2, \bar{5}) \,
t_1(k, \bar{4}; 3, \bar{6}) \nonumber \\
&\hs*{5mm}
 +  t_1(i, \bar{4}; 1, \bar{6}) \,  t_1(j, \bar{6}; 2, \bar{5}) \, t_1(k, \bar{5}; 3, \bar{4}) \Big]S(\bar{4}) \, S(\bar{5}) \, S(\bar{6})  \,,  
\label{bb3}
\end{align}
shown graphically in Fig.~\ref{fig:tb}. Here $\epsilon_{ijk}$ is the usual antisymmetric tensor, and a sum over
$i, j, k = 1, 2, 3$ is implied. 
The amplitude Eq.~(\ref{bb3}) is totally symmetric in $(123)$, and antisymmetric with respect to interchanges of two 4-momenta 
in the final state.
    
%===============================================================================
%===============================================================================

The diagrams shown in Fig.~\ref{fig:tb} are the three-particle analogues of the two-particle amplitude
shown in Fig.~\ref{fig:t2}.  
We call these class B terms ``medium induced processes'', because in the case of energy-independent vertices, like the
bare interaction or any energy-independent approximation to the ladder $t$ -matrix, at least one of the intermediate 
lines in Fig.~\ref{fig:tb} must be a hole line.
The intermediate states in these processes involve, then, in addition to the three given particles also a particle-hole pair, i.e., the class B terms
actually describe 4-particle processes. Those terms do not show up in the usual Faddeev series,
because there it is assumed from the outset that the first and the last interactions
occur among the three given particles, and not between one of them and a background particle.

Finally, all class C terms come from $\delta K_3^{(2)}/{\delta S(3)}$, and arise by opening the pairs 
$(\bar{2}, \bar{3})$, $(\bar{2}, \bar{6})$, $(\bar{3}, \bar{6})$, $(\bar{4}, \bar{5})$, $(\bar{4}, \bar{6})$, $(\bar{5}, \bar{6})$
in the self energy $\Sigma_3$ of Fig.~\ref{fig:se}\textcolor{blue}{b}. 
As already mentioned, they are obtained by replacing one of the vertices $t_1$
in Eq.~(\ref{k32}) by $t_2$ of Eq.~(\ref{t2nf}). The class C terms are then simply expressed as 
\begin{align}
\tilde{h}_3^{(\rm C)}(1,2,3) &= - \Big( t_1(1, 2; 3, \bar{4}) \, t_2(3, \bar{4}; 1, 2)\nonumber \\ 
& \hs*{10mm} + t_2(1,2; 3, \bar{4}) \, t_1(3, \bar{4}; 1, 2) \Big) \, S(\bar{4}) \nonumber \\
&+ \left(1 \leftrightarrow 3\right) + \left(2 \leftrightarrow 3\right)\,.
\label{typec}
\end{align}

To summarize, we have derived the formulas up to order $t_1^3$ for the terms in the three-particle amplitude
$\tilde{h}$, i.e., for the terms given in Eqs.~(\ref{hone}), (\ref{htwo}) and (\ref{hfive}).  
%which do not vanish when retardation effects in the two-particle $t$-matrix are neglected, i.e., which could be 
%assessed by using an energy-indenpendent approximation. 
Among them, we recovered the first two terms in the Faddeev 
series shown in Figs.~\ref{fig:sample} and \ref{fig:ta}, and associated medium induced interactions 
shown in Fig.~\ref{fig:tb}.
The latter ones describe the interactions of the three given particles with the particles in the Fermi sea. 
We have also confirmed that the resulting three-particle amplitude satisfies the Pauli principle.  

We note that the expansion of the $t$-matrix and the self energy in powers of $t_1$, given by 
Eqs.~(\ref{tsplit}) and (\ref{sesplit}), can also be used to expand the product terms of Eq.~(\ref{hprod}). 
As we mentioned earlier, those product terms are associated with self energy subgraphs in
the graphs for $\Sigma$, i.e., in analogy to the $Z$-factors in the two-particle amplitude~\cite{Negele:1988aa} 
they could be represented graphically by cutting simultaneously a line in a 
self energy subgraph and the line to which this subgraph is attached. Such graphical representations, however,
are neither illuminating nor useful, and we find it more convenient to use the expression Eq.~(\ref{hprod}) without
associating diagrams with it. 

Returning finally to the notation of Eq.(\ref{split}) of the previous section, we have shown that the 2pc and 3pc pieces
are given to third order in the ladder $t$-matrix by
\begin{align}
h^{(\rm 2pc)} &= \tilde{h}_2 + \tilde{h}_3^{(C)}  \,,  \nonumber \\
h^{(\rm 3pc)} &= \tilde{h}_3^{(A)} + \tilde{h}_3^{(B)} \,.    \nonumber   
\end{align}

%===============================================================================
%===============================================================================
\section{SUMMARY AND FINAL REMARKS\label{sec:IV}}
The motivation for the first part of our present work was the rapidly expanding interest in the symmetry energy
of nuclear matter ($a_s$) and its slope parameter ($L$), two physical quantities which have decisive impact on the 
structure of nuclei and neutron stars. In view of the many model calculations based on effective interactions,
our primary aim was to discuss these quantities in the model independent framework of the Fermi liquid theory of
Landau and Migdal. 
The main result is summarized by Eq.~(\ref{result}), which is exact and remarkably
simple, because it does not involve any momentum derivatives of the effective mass or interaction
parameters. The physically most interesting part of this relation is the isovector three-particle $s$-wave
Landau-Migdal parameter $H_0'$. We estimated the two-particle correlation contribution to this term, 
represented by Fig.~\ref{fig:sample}\textcolor{blue}{a}, and found that it gives a moderate contribution of roughly
$20 \sim 30\%$ of the leading term [$C_0$ of Eq.~(\ref{c0})]. The leading term alone is within the empirical
limits given by Eq.~(\ref{lim}), if the parameter $\mu$ of Eq.~(\ref{mu}) is non-zero and positive.
From our simple estimates, we found that the effect of $\mu$, which reflects the proton-neutron
mass difference in isospin asymmetric matter, and of the three-particle interaction term
$H_0'$, work in the same direction and are of similar magnitude. If one could assess these two
quantities more quantitatively by model calculations or other empirical information, our result
will be useful to further pin down the slope parameter $L$ and the associated
symmetry pressure, which plays an important role in nuclei and compact stars.   

Because one may expect that there exist several more relations between
three-particle interaction parameters and observables, it is desirable to have more understanding
about the physics of the three-particle interaction term introduced in Eq.~(\ref{vare}). This was the motivation 
for the second part of our present work. For this purpose,
we extended the well known discussions on the two-particle amplitude in the Fermi liquid theory to the three-particle case.
Because, to our knowledge, such a discussion has not yet been presented in the literature, we limited
ourselves to the case of symmetric nuclear matter. 
The general result for the three-particle amplitude, which is shown in Eqs.~(\ref{hsplit})--(\ref{hfive}), 
involves the three-particle
kernel of Eq.~(\ref{k3}). We specified its form by using the ladder approximation, thereby
making contact to Bethe-Brueckner-Goldstone (BBG) theory. Besides the first few terms of the in-medium Faddeev series,
we found a class of medium induced processes of the same order, which have their origin in the interaction
between the three given particles at the Fermi surface and the particles in the Fermi sea. 
We derived the basic formulas for those processes, but detailed model calculations are necessary to
assess their role in a quantitative way. We have also outlined the way to extend our method to higher
orders in the basic ladder $t$-matrix, and it would be very interesting to see which kind of medium induced
three-body processes appear in higher orders, in addition to the well known Faddeev-type processes.

We finally add a few remarks on the relation between our approach to other methods mentioned in Sec.~\ref{sec:I}. 
First, one basic point of our approach is the expansion of the self energy ($\Sigma$) in terms of skeleton diagrams, which
by definition do not contain self energy subgraphs. The skeletons of $\Sigma$ can be considered as functionals of the
full single particle propagators, including both particle and hole parts. 
The effects of self energy subgraphs are thus separated from the start (Eq.~(\ref{hsplit})), and need not be considered explicitly in the
calculation of the three-particle kernel (Eq.~(\ref{k3})).
These points have much in common with the self-consistent Green's function method~\cite{Dickhoff:2004xx,Carbone:2013eqa}.
Such an approach is, however, not possible for the energy density ($E$), because one cannot define a skeleton of 
$E$~\cite{Nozieres:1964aa}.
On the other hand, the calculations done in the BBG theory~\cite{Bethe:1965zz,Day:1972zz,Day:1978tn} -- and subsequent important extensions to 
variational calculations and inclusion of ring diagrams~\cite{Day:1978zz,Day:1981zz,Day:1985jn} --
expand the energy density in the number of
hole lines, and the subset of graphs with three hole lines corresponds to the Faddeev series with internal
particle propagators only. The motivation for this approach is the low-density expansion of $E$~\cite{Bethe:1965zz},
where -- in our notations -- the three functional derivatives in  
$h = \delta^3 E/(\delta n^3)$ (see Eq.~(\ref{start}))
can be considered to act on the hole lines only. Higher order terms in the hole-line expansion, which have been
considered explicitly in Ref.~\cite{PhysRev.187.1269}, correspond to higher order terms in the low-density expansion of $E$.

Because in the present paper we did not derive the full Faddeev series including the associated medium induced 
processes, it is not yet possible to assess the advantage of one method over the other. 
%However, if our
%Fermi liquid approach can be properly extended, we expect to have it the following advantages: First,
%the effects of self energy subgraphs are clearly separated from the start and need not be considered
%explicitly in the calculation of the three-particle kernel. Second, it avoids the distinction between particle and
%hole parts of the propagators, leading to a more compact formulation. 
As discussed already in Sec.~\ref{sec:I}, the major advantage of the Fermi liquid approach 
is to establish model independent relations between parameters characterizing the interactions between two quasiparticles
at the Fermi surface and observable quantities. In a previous paper~\cite{Bentz:2019lqu} and in the
present work, we have shown that such model independent relations exist also between three-particle interaction
parameters and other observable quantities. We therefore believe that it is worth while to work on an extension
of our present approach in the following directions: First, as outlined in Sec.~\ref{sec:IIIc}, to reproduce the full Faddeev 
series and associated medium induced three-particle interactions; second, to search for new relations between 
three-particle interaction parameters and observables; and third to assess the three-particle interaction parameters
quantitatively in model calculations. 

A second and final comment concerns the question whether it is necessary at all to consider the three-particle term
explicitly, as we have done from the outset in Eq.~(\ref{vare}). 
%or whether its effects can be renormalized into
%the two-body interaction. 
It is actually well known, and can also be inferred from the structure of Eq.~(\ref{vare}), that
the effects of the three-particle term can be renormalized into the two-particle interaction, but different renormalization
conditions lead to different coefficients for the three-particle term~\cite{(HeYeTongLang):2013dsa}. For example, 
Ref.~\cite{Vautherin:1971aw} employed the condition that 
the renormalized two-particle interaction gives the same total energy as the original interaction including the 
three-particle term explicitly.
This is different from the procedure where one imposes the condition that the renormalized two-particle interaction gives the same 
single-particle energies, or from the method where a naive average 
over one of the three particles is used to define
a renormalized two-particle interaction~\cite{(HeYeTongLang):2013dsa}. 
To our opinion, the main motivation for considering the three-particle term explicitly is that the associated Landau-Migdal 
parameters enter into simple, exact, and
model independent relations to other physical quantities. We have demonstrated this in a previous paper for the
skewness, and in the present work for the slope of the symmetry energy of nuclear matter. We hope that
extensions of our framework will lead to several other useful relations of this kind.

%===============================================================================
%===============================================================================
\begin{acknowledgments}
The authors thank K. Noro (Tokai Univeristy) for technical help with the figures.
W.B. thanks the staff of the Theory Group at Argonne National Laboratory for their kind arrangements to visit the Lab 
in March 2020, and to late Prof. A. Arima for his continuous encouragement and support to study problems of nuclear structure physics.    
The work of I.C. was supported by the U.S. Department of Energy, Office of Science, Office of Nuclear Physics, 
contract no. DE-AC02-06CH11357.
\end{acknowledgments}

%================================================================================
%==================================================================================
\appendix
\section{GALILEI INVARIANCE FOR ISOSPIN ASYMMETRIC MATTER\label{app:A}}
In this Appendix we wish to derive an exact relation for the in-medium proton-neutron mass difference from
Galilei invariance, and compare it with the approximate relation Eq.~(\ref{muo}) which has often been assumed in the literature~\cite{Goriely:2010bm,Li:2018lpy}.

As usual, one considers the variation of the quasiparticle energy
which arises from the change of the distribution function due to a Galilei transformation
from the rest system of nuclear matter to a system which moves with velocity
$\vect{u} \equiv \vect{q}/M$, where $M$ is the free nucleon mass. 
To first order in $q$ these variations are given by 
\begin{align}
\delta n^{(\tau)}_{\vect{k}} &= 
%\theta \left(p_F - |\vect{k} + \vect{p}| \right)  - \theta(p_F - k)
 - (\vect{\hat k} \cdot \vect{q}) \, \delta(p^{(\tau)} - k) \,,
\label{gal1} \\
\delta \varepsilon^{(\tau_1)}(\vect{k}; \{\rho\}) &=  2 \int \frac{{\rm d}^3 k_2}{(2 \pi)^3} \, 
f^{(\tau_1 \tau_2)}(\vect{k}, \vect{k}_2; \{\rho\}) \, \delta n^{(\tau_2)}_{\vect{k}_2}   \nonumber \\
&\hs*{-10mm}
= - \frac{1}{3 \pi^2} \left(\vect{\hat k} \cdot \vect{q}\right) \, f_1^{(\tau_1 \tau_2)}(k, p^{(\tau_2)}; \{\rho\}) \, 
p^{(\tau_2) 2} \,,
\label{gala} 
\end{align}
where we use the notations introduced in Eq.~(\ref{vare}).
On the other hand, the quasiparticle energy should transform in the same way as a Hamiltonian
in classical mechanics, i.e;
$\varepsilon^{' (\tau)}(\vect{k}';\{\rho\}) = \varepsilon^{(\tau)}(\vect{k};\{\rho\}) - \frac{\vect{k}\cdot\vect{q}}{M}
+ \frac{q^2}{2M}$, where $\vect{k}' = \vect{k} - \vect{q}$. From this it follows that
$\varepsilon^{' (\tau)}(\vect{k};\{\rho\}) = \varepsilon^{(\tau)}(\vect{k} + \vect{q}; \{\rho\}) - 
\frac{\vect{k}\cdot \vect{q}}{M} - \frac{q^2}{2M}$, and to first order in $q$, 
\begin{align}
\delta \varepsilon^{(\tau)}(\vect{k}; \{\rho\})   
= \left(\vect{k} \cdot \vect{q} \right) \left(\frac{1}{M^{* (\tau)}(k,\{\rho\})} - \frac{1}{M} \right)\,,
\label{varg}
\end{align}
where we used the usual definition of the effective mass in terms of the quasiparticle velocity. 
The requirement that Eqs.~\eqref{gala} and \eqref{varg} are identical leads to the relations
\begin{align}
\label{mstp}
&\frac{k}{M^{*(p)}(k;\{\rho\})} + \frac{1}{3 \pi^2} \nonumber \\
&\hs*{4mm}\times \left[f_1^{(pp)}(k, p^{(p)}; \{\rho\}) \, p^{(p)2} 
+ f_1^{(pn)}(k, p^{(n)}; \{\rho\}) \, p^{(n)2} \right] \nonumber \\
&= \frac{k}{M} \,, \\[0.5em]
\label{mstn}  
&\frac{k}{M^{*(n)}(k;\{\rho\})} + \frac{1}{3 \pi^2} \nonumber \\
&\hs*{4mm}\times \left[f_1^{(np)}(k, p^{(p)}; \{\rho\}) \, p^{(p)2} 
+ f_1^{(nn)}(k, p^{(n)}; \{\rho\}) \, p^{(n)2} \right] \nonumber \\
&= \frac{k}{M} \,,  
\end{align}
which  hold for any values of $k$ and background densities $\{\rho\} = \{\rho^{(p)}, \rho^{(n)}\}$. 
For the case $k=p^{(p)}$ in Eq.~(\ref{mstp})
and $k=p^{(n)}$ in Eq.~(\ref{mstn}), these are the familiar effective mass relations in asymmetric nuclear matter, derived 
first in Ref.~\cite{Sjoberg:1976tq}. 
The sum of Eqs.~(\ref{mstp}) and (\ref{mstn}) in the isospin symmetric limit gives
\begin{align}
\frac{k}{M^*(k; \rho)} + \frac{2 p^2}{3 \pi^2} \, f_1(k,p; \rho) = \frac{k}{M} \,,
\label{sum}
\end{align}
where $f_1 \equiv \left(f_1^{(p)} + f_1^{(n)}\right)/2$, and $\rho = 2 p^3/(3 \pi^2)$ is the total baryon density 
with $p$ the corresponding Fermi momentum in the isospin symmetric limit. For $k=p$, this becomes the familiar 
Landau effective mass relation 
\begin{align}
\frac{M^*}{M} = 1 + \frac{F_1}{3} \,,    \label{landau}
\end{align}
where the dimensionless parameter $F_1$ is defined as usual~\cite{Negele:1988aa}.

The difference of Eq.~(\ref{mstp}) and (\ref{mstn}) at fixed $k=p$ is
\begin{align}
&\frac{1}{M^{*(p)}(p;\{\rho\})} - \frac{1}{M^{*(n)}(p;\{\rho\})}
= \frac{-1}{3 \pi^2 p} \nonumber \\ 
&\times
\left[ \left( f_1^{(pp)}(p, p^{(p)}; \{\rho\}) -  f_1^{(np)}(p, p^{(p)}; \{\rho\}) \right) \, p^{(p)2}
\right. 
\nonumber \\
& \left. -   \left( f_1^{(nn)}(p, p^{(n)}; \{\rho\}) -  f_1^{(pn)}(p, p^{(n)}; \{\rho\}) \right) \, p^{(n)2} \right] \,.
\label{diff}
\end{align}
We wish to consider the terms of first order in $\rho^{(3)}$ of Eq.~(\ref{diff}), and then take the isospin symmetric limit. 
For this purpose, we use [see Eq.~(\ref{rpn})]
\begin{align}
\rho^{(p)} = \frac{\rho}{2} + \frac{\rho^{(3)}}{2} \,, \,\,\,\,\,\,\,\,\,\,\,\,
\rho^{(n)} = \frac{\rho}{2} - \frac{\rho^{(3)}}{2} \,,  
\label{rho}
\end{align}
as well as the corresponding relations for the Fermi momenta
\begin{align}
p^{(p)} = p + p^{(3)} \,, \,\,\,\,\,\,\,\,\,\,\,\,
p^{(n)} = p - p^{(3)} \,,     \label{fermi}
\end{align}
where the first order relation between $\rho^{(3)}$ and $p^{(3)}$ is given by
$\rho^{(3)} = \frac{2 p^2}{\pi^2} \, p^{(3)}$. 

The l.h.s. of Eq.~(\ref{diff}), to first order in $\rho^{(3)}$, is given by [see Eq.~(\ref{biv})]
\begin{align}
\frac{1}{M^{*(p)}(p;\{\rho\})} - \frac{1}{M^{*(n)}(p;\{\rho\})} = \frac{\rho^{(3)}}{p} \,
\frac{\partial f_0'}{\partial p} \,. 
\label{lhs}
\end{align}
On the r.h.s. of Eq.~(\ref{diff}), we expand all quantities about the isospin symmetric limit,
i.e., about the Fermi momentum $p$ and the background density $\rho$, using Eqs.~(\ref{rho}) and (\ref{fermi}). For example, for the first term in the second line of Eq.~(\ref{diff}) we write, up to first order in $\rho^{(3)}$:
\begin{align}
f_1^{(pp)}(p, p^{(p)}; \{\rho\}) &= f_1^{(pp)} + \frac{\pi^2}{2 p^2} 
\frac{\partial f_1^{(pp)}(p, k_2)}{\partial k_2}|_{k_2=p}   \nonumber \\  
&+ \frac{\rho^{(3)}}{2} \left(h_1^{(ppp)} - h_1^{(ppn)} \right) \,.
\nonumber
\end{align}
In this way we obtain for the r.h.s. of Eq.~(\ref{diff}), to first order in $\rho^{(3)}$ and in the isospin symmetric limit 
\begin{align}
\frac{\rho^{(3)}}{p} \left( - \frac{4}{3p} f_1' - \frac{1}{3} \frac{\partial f_1'}{\partial p} 
- \frac{4 p^2}{3 \pi^2} h_1' \right) \,. \label{rhs}
\end{align}
Here $h_1'$ is defined by
\begin{align}
h_1' \equiv \frac{\delta f_1}{\delta \rho^{(3)}} = \frac{1}{4} 
\left(h_1^{(ppp)} - h_1^{(ppn)} + h_1^{(pnp)} - h_1^{(pnn)} \right) \,,
\label{h1p}
\end{align}
where the amplitudes $h_1^{(\tau_1 \tau_2 \tau_3)}$ were defined in Eq.~(\ref{old4}). 
Comparison of Eqs.~(\ref{lhs}) and (\ref{rhs}) then gives the identity
\begin{align}
p \,\frac{\partial}{\partial p} \left(f'_0 + \frac{1}{3} f'_1\right) + \frac{4}{3} f'_1 
+ \frac{4 p^3}{3 \pi^2} \, h'_1 = 0 \,,
\label{gal2}
\end{align}
which is simply obtained from its isoscalar counterpart, Eq.~(\ref{old5a}), by attaching a prime to
all quantities.\footnote{As for the case of its isoscalar counterpart, the relation given in Eq.~(\ref{gal2}) can also be derived more directly from the Galilei invariance of the isovector two-particle scattering amplitude, although we do not go into details here.}
Using this identity to eliminate the derivative of $f_0'$ in Eq.~(\ref{lhs}), we obtain finally 
\begin{align}
&\frac{1}{M^{*(p)}} - \frac{1}{M^{*(n)}} = \frac{\rho^{(3)}}{p^2} \left( p \frac{\partial f_0'}{\partial p}
- \frac{\pi^2}{M^{*2}} \frac{\partial M^*}{\partial p} \right)   \nonumber \\
& = - \beta \frac{2 p}{3 \pi^2} \left[  \frac{4}{3} f_1' + \frac{p}{3}  \frac{\partial f_1'}{\partial p}
+ \frac{\pi^2}{M^{*2}} \frac{\partial M^*}{\partial p} + \frac{4 p^3}{3 \pi^2} \, h'_1 \right] \,, 
\label{finapp}
\end{align}
where we introduced the asymmetry parameter ${\displaystyle \beta = \frac{\rho^{(3)}}{\rho} = \frac{Z-N}{A}}$.

The relation given in Eq.~(\ref{finapp}) can be used to express $\mu$ of Eq.~(\ref{mu}) by the interaction parameters.
In terms of the dimensionless parameters used in the main text 
($F_1' = (2pM^*/\pi^2) \, f_1'$ and $H_1' = (4 p^4 M^*/3 \pi^4) h_1'$), we obtain\footnote{The derivative of $F_1'$ in Eq.~(\ref{muc}) by definition acts only on $f_1'$, and not on the defining
prefactor $2pM^*/\pi^2$.}
\begin{align}
\mu &= \frac{2}{3} \frac{M^*}{M} \left( \frac{2}{3} F_1' + \frac{p}{6} \frac{\partial F_1'}{\partial p}
+ \frac{p}{M^*} \frac{\partial M^*}{\partial p} + H_1' \right) \,.
\label{muc}
\end{align}
Eq.~(\ref{muc}) is a rather complicated expression and not very useful in practice, therefore we avoided it in the main text. It is different from the simple relation of Eq.~(\ref{muo}), which has been found to be approximately valid
in model calculations based on Skyrme-type interactions~\cite{Goriely:2010bm,Li:2018lpy}.

Another way to express the result of Eq.~(\ref{muc}) is via an ``isovector effective mass'' $\tilde{M}_V^*$, which is defined
by
\begin{align}
\frac{1}{M^{*(p)}} - \frac{1}{M^{*(n)}} \equiv 2 \beta \left(\frac{1}{M^*} - \frac{1}{\tilde{M}_V^*} \right) \,.
\label{defmv}
\end{align}
By using the relation $\mu = 2 \frac{M^*}{M} \left( \frac{M^*}{\tilde{M}_V^*} - 1 \right)$, we can express Eq.~(\ref{muc}) as
\begin{align}
\frac{M^*}{\tilde{M}_V^*} &= 1 + \frac{2}{9} F_1' + \frac{1}{18} p \frac{\partial F_1'}{\partial p} + \frac{p}{3 M^*}
\frac{\partial M^*}{\partial p} + \frac{1}{3} H_1' \,,
\label{mv}
\end{align}
Again, this is more complicated than the simple relation $\frac{M^*}{\tilde{M}_V^*} = 1 + F_1'/3$, which was found to
be valid in model calculations using Skyrme-type interactions.

We note that the more standard definition of ``isovector effective mass'' ($M_V^*$) is 
via the enhancement factor $(1 + \kappa)$ of the electric dipole (Thomas-Reiche-Kuhn) sum rule value~\cite{Hutt:1999pz},
or via the isovector combination of orbital angular momentum $g$-factors~\cite{Arima:1973rzn}:
\begin{align}
\frac{M}{M_V^*} = 1 + \kappa \simeq g_{\ell}^{(p)} - g_{\ell}^{(n)}.
\nonumber
\end{align}
This quantity $M_V^*$ is related to $F_1'$ by~\cite{Migdal:1967aa,Bentz:2003zb} 
\begin{align}
\frac{M^*}{M_V^*} = 1 + \frac{F_1'}{3}  \,.
\label{miv} 
\end{align}
Comparing with Eq.~(\ref{mv}), we see that the quantities $\tilde{M}_V^*$ and $M_V^*$ are generally different,
although numerically they seem to be of similar magnitude in calculations using Skyrme-type interactions.

\section{DERIVATION OF RELATIONS USED IN SEC.~\ref{sec:III}\label{app:B}}

\subsection{Proof of Eqs.~(\ref{dsdn}) and (\ref{ds2dn})}
Here we show that the first term in Eq.~(\ref{dsdn}) is obtained if the functional derivative  
$\delta/\delta n_{\vect{k}'}$ acts only on the last term in the denominator of the propagator given in Eq.~(\ref{s}).
For this purpose, let us define $A(k) \equiv k_0 - \varepsilon_0(\vect{k}) - \Sigma(k)$, and consider
the contribution from the functional derivative acting only on the term $-i \eta (2 n_{\vect{k}} - 1)$
in the denominator: 
\begin{align}
&  \frac{\delta}{\delta n_{\vect{k}'}} \frac{1}{A(k) - i \eta \left(2 n_{\vect{k}} - 1 \right)}
= \frac{\delta}{\delta n_{\vect{k}'}} \frac{i \eta \left(2 n_{\vect{k}} - 1 \right)}{A^2 + \eta^2} 
\nonumber \\
&\hs*{18mm} = 2 \pi i \, \delta(A(k))\ \frac{ \delta n_{\vect{k}}}{\delta n_{\vect{k}'}}, \no \\
&\hs*{18mm} = i \, \left(2 \pi\right)^4 Z_{\vect{k}} \delta\left(k_0 - \varepsilon(\vect{k})\right)  
\delta^{(3)}(\vect{k} - \vect{k}').
\end{align} 
This gives the first term of Eq.~(\ref{dsdn}). The second term of Eq.~(\ref{dsdn}) is obtained if the functional
derivative $\delta/\delta n_{\vect{k}'}$ acts on the self energy $\Sigma(k)$.

In order to show Eq.~(\ref{ds2dn}), we add an auxiliary infinitesimal constant ($\alpha$) to the self energy
in the denominator of Eq.~(\ref{s}). If we call this new propagator $\tilde{S}(k)$, then obviously up to order $\alpha$
$\tilde{S}(k) = S(k) + \alpha S^2(k)$. Therefore, to show Eq.~(\ref{ds2dn}), we have to take the term of order $\alpha$
of the following expression:
\begin{align}
\frac{\delta \tilde S(k)}{\delta n_{\vect{k}'}} &= 
i \left(2\pi \right)^4 \delta^{(3)}\left(\vect{k}-\vect{k}'\right) 
\,\delta\left(k_0 - \varepsilon(\vect{k}) - \alpha \right)\, \tilde{Z}_{\vect{k}}  \nonumber \\
& + \tilde{S}^2(k) \, \frac{\delta \Sigma(k)}{\delta n_{\vect{k}'}},
\label{z}
\end{align}
where we used the identity given in Eq.~(\ref{dsdn}) with  
$\tilde{Z}_{\vect{k}} = (1 - \Sigma'(k_0 = \varepsilon(\vect{k}) + \alpha))^{-1}$. 
Expanding Eq.~(\ref{z}) about $\alpha=0$ and
taking the term of order $\alpha$ immediately leads to Eq.~(\ref{ds2dn}).

\subsection{Proof of Eqs.~(\ref{hsplit})  -  (\ref{hfive})}
Applying $\frac{\delta}{\delta n_{\vect{k}_3}}$ to each term in the BS equation of Eq.~(\ref{bs1}) we obtain,
using Eqs.~(\ref{dk2dn}), (\ref{ds2dn}) and (\ref{deft3})
\begin{align}
& t^{(3)}(k_1, k_2, k_3) = A(k_1, k_2, k_3) + B(k_1, k_2, k_3) \nonumber \\
& \hs*{5mm} - i \int \frac{{\rm d}^4 k}{(2\pi)^4} \, K^{(2)}(k_1, k) \, S^2(k) \, t^{(3)}(k, k_2, k_3) \,,  
\label{bs2}
\end{align}
where we split the driving term into two parts $A$ and $B$, which are defined as
\begin{align}
& A(k_1, k_2, k_3) = \left(\frac{\partial}{\partial k_{30}} + Z_{\vect{k}_3} \, \Sigma''(\vect{k}_3)\right) 
K^{(2)}(k_1, k_3) \, t(k_3, k_2),
\label{a}   \\
& B(k_1, k_2, k_3) = K^{(3)}(k_1, k_2, k_3)   \label{b1} \\
& \hs*{5mm} -i   \int \frac{{\rm d}^4 k}{(2\pi)^4} \, K^{(3)}(k_1, k_2, k) \, S^2(k) \, t(k, k_3) 
\label{b2} \\
& \hs*{5mm} - i  \int \frac{{\rm d}^4 k}{(2\pi)^4} \, K^{(3)}(k_1, k, k_3) \, S^2(k) \, t(k, k_2)  
\label{b3} \\
& \hs*{5mm} -  \int \frac{{\rm d}^4 k}{(2\pi)^4}  \int \frac{{\rm d}^4 k'}{(2\pi)^4} \, K^{(3)}(k_1, k, k') \, 
S^2(k) \, t(k, k_3)  \nonumber \\ 
& \hs*{10mm} \times S^2(k') \, t(k', k_2)  \label{b4} \\
& \hs*{5mm} - 2 i  \int \frac{{\rm d}^4 k}{(2\pi)^4} K^{(2)}(k_1, k) \, S^3(k) \, t(k, k_2) \, t(k, k_3) \,. 
\label{b5}
\end{align}
Here the term $A$ and the last term of $B$ arise from the functional derivative of $S^2(k)$ in Eq.~(\ref{bs1}),
by using Eqs.~(\ref{ds2dn}) and (\ref{deft}), the first two terms of $B$ come from the functional derivative of the
driving term ($K^{(2)}$) in Eq.~(\ref{bs1}), by using Eq.~(\ref{dk2dn}), and the third and the fourth terms of $B$ come from
from the functional derivative of $K^{(2)}$ under the integral in Eq.~(\ref{bs1}), by using Eq.~(\ref{dk2dn}).    

At first sight, Eq.~(\ref{bs2}) may look like a complicated integral equation, but actually this is not the
case: The kernel of this integral equation is the same as in the basic BS equation of Eq.~(\ref{bs1}), and
therefore Eq.~(\ref{bs2}) can easily be resolved in the following way: 

We first note that Eq.~(\ref{bs2}) can be expressed as two separate
integral equations, i.e.;
\begin{align}
t^{(3)}(k_1, k_2, k_3) = X (k_1, k_2, k_3) + Y(k_1, k_2, k_3),
\label{xy}
\end{align}
where $X$ and $Y$ are solutions of the two separate equations
\begin{align}
X(k_1, k_2, k_3) &= A(k_1, k_2, k_3) \nonumber \\
& \hs*{-10mm} - i \int \frac{{\rm d}^4 k}{(2\pi)^4} \,
K^{(2)}(k_1, k) S^2 (k) X(k, k_2, k_3),   \label{x} \\
Y(k_1, k_2, k_3) &= B(k_1, k_2, k_3)      \nonumber \\
& \hs*{-10mm} -i \int \frac{{\rm d}^4 k}{(2\pi)^4} \,
K^{(2)}(k_1, k) S^2 (k) Y(k, k_2, k_3) \,.   \label{y} 
\end{align}
Let us first consider Eq.~(\ref{x}). Inserting here first the term 
$K^{(2)}(k_1, k_3) \, \left(\partial t(k_3, k_2)/\partial k_{30}\right)$ in the expression (\ref{a}) for $A$
and using the equation Eq.~(\ref{bs1}) for the two-particle t-matrix, it is clear that this
terms gives a contribution $t(k_1, k_3) \, \left(\partial t(k_3, k_2)/\partial k_{30}\right)$ to $X$.
Next, inserting the term $\left(\partial K^{(2)}(k_1, k_3)/\partial k_{30}\right) \, t(k_3, k_2)$
in the expression Eq.~(\ref{a}) for $A$ and using the partial derivative of Eq.~(\ref{bs1}) w.r.t.
$k_{20}$
\begin{align}
\frac{ \partial t(k_1, k_2)}{\partial k_{20}}  &= \frac{ \partial K^{(2)}(k_1, k_2)}{\partial k_{20}}  \nonumber \\
& \hs*{-5mm} - i \int \frac{{\rm d}^4 k}{(2\pi)^4} \, K^{(2)}(k_1, k) \, S^2(k) \, 
\frac{ \partial t(k, k_2)}{\partial k_{20}}  \,,
\label{bs1p}
\end{align}
we see that this term gives a contribution $\left(\partial t(k_1, k_3)/ \partial k_{30}\right) \, 
t(k_3, k_2)$ to $X$. Finally, inserting the term $\left(Z_{\vect{k}_3} \, \Sigma''(\vect{k}_3)\right) 
\left(K^{(2)}(k_1, k_3) \, t(k_3, k_2)\right)$ in the expression given in Eq.~(\ref{a}) for $A$ and using Eq.~(\ref{bs1}) for the $t$-matrix shows that this term gives a contribution
$\left(Z_{\vect{k}_3} \, \Sigma''(\vect{k}_3)\right) 
\left(t(k_1, k_3) \, t(k_3, k_2)\right)$ to $X$. As a result, $X(k_1, k_2, k_3)$ is obtained as
\begin{align}
X(k_1, k_2, k_3) &= \frac{\partial}{\partial k_{30}} \left(t(k_1, k_3)  \, t(k_3, k_2) \right)  \nonumber \\
& + \left(Z_{\vect{k}_3} \, \Sigma''(\vect{k}_3) \right) \, t(k_1, k_3) \, t(k_3, k_2) \,.
\label{xsol}
\end{align}
After multiplying the $Z$-factors of the three particles, according to Eqs.~(\ref{deft3}) and (\ref{ff}),
and going to the Fermi surface, this gives a contribution
\begin{align}
&\frac{1}{2} \left( \frac{\partial f(\vect{k}_1, \vect{k}_3)}{\partial \varepsilon} 
f(\vect{k}_2, \vect{k}_3) + \frac{\partial f(\vect{k}_2, \vect{k}_3)}{\partial \varepsilon}
f(\vect{k}_1, \vect{k}_3 \right) \nonumber \\
&+ \left(Z \, \Sigma'' \right) \, f(\vect{k}_1, \vect{k}_3) \, f(\vect{k}_2, \vect{k}_3) \,, 
\label{part1}
\end{align}
to the three-particle amplitude $h(\vect{k}_1, \vect{k}_2, \vect{k}_3)$, where we used the definitions explained in Eq.~(\ref{depsnew}) and in the text below that equation.  
Adding Eq.~(\ref{part1}) to Eq.~(\ref{part}) of the main text gives the totally symmetric product part
$h^{({\rm prod})}(\vect{k}_1, \vect{k}_2, \vect{k}_3)$, as given by Eq.~(\ref{hprod}).

Next we consider the quantity $Y(k_1, k_2, k_3)$ of Eq.~(\ref{y}), which is identical to 
$\tilde{h}(k_1, k_2, k_3)$ in the main text.
According to Eqs.~(\ref{b1})--(\ref{b5}), the function 
$B(k_1, k_2, k_3)$ splits into five pieces, so $B(k_1, k_2, k_3) = \sum_{i=1}^{5} B_i(k_1, k_2, k_3)$.
Therefore also $Y(k_1, k_2, k_3)$ splits into five pieces, $Y(k_1, k_2, k_3) = \sum_{i=1}^{5} Y_i(k_1, k_2, k_3)$,
where each $Y_i$ satisfies the equation
\begin{align}
Y_i(k_1, k_2, k_3) &= B_i(k_1, k_2, k_3) \nonumber \\
&\hs*{-10mm}
- i \int \frac{{\rm d}^4 k}{(2\pi)^4} K^{(2)}(k_1, k) \,S^2(k) \, Y_i(k, k_2, k_3).
\end{align}
Iteration of this equation, and comparison with Eq.~(\ref{bs1}) shows that
\begin{align}
Y_i(k_1, k_2, k_3) &= B_i(k_1, k_2, k_3) \nonumber \\
&- i \int \frac{{\rm d}^4 k}{(2\pi)^4} t(k_1, k) \,S^2(k) \, B_i(k, k_2, k_3) \nonumber  \\
& \equiv B_i(k_1, k_2, k_3) + \tilde{B}_i(k_1, k_2, k_3) \,.   \label{yi}
\end{align} 
Then the sum $Y = \sum_{i=1}^5 Y_i= \sum_{i=1}^5 \left(B_i + \tilde{B}_i \right)$ is identical to
Eqs.~(\ref{hone})--(\ref{hfive}) of the main text. In more detail,
\begin{itemize}
\item the term given in Eq.~(\ref{hone}) is identical to $B_1$;
\item the term given in Eq.~(\ref{htwo}) is identical to the sum $\left(B_2 + B_3 + \tilde{B}_1 \right)$;
\item the term given in Eq.~(\ref{hthree}) is identical to the sum $\left(B_4 + \tilde{B}_2 + \tilde{B}_3 \right)$;
\item the term given in Eq.~(\ref{hfour}) is identical to $\tilde{B}_4$; and
\item the term given in Eq.~(\ref{hfive}) is identical to $\left( B_5 + \tilde{B}_5\right)$.
\end{itemize}

\subsection{Two-body $t$-matrix in third order and self energy in fourth order}
The two-body $t$-matrix in third order of the ladder $t$-matrix $t_1$ is given by Eq.~(\ref{first})--(\ref{third}) in the
main text. 
Inserting the off-forward forms of $K_2^{(2)}$ and $K_3^{(2)}$, which are obtained from Eqs.~(\ref{k22}) and (\ref{k23}) 
by replacing $1 \rightarrow 1'$, $2 \rightarrow 2'$ in the final states (first two arguments) of each $t$-matrix,
and using $t_2$ as given by Eq.~(\ref{t2nf}), we obtain
\begin{align}
&t_3(1', 2'; 1, 2) = \left[t_1(1', \bar{5}; \bar{6}, \bar{4}) \, t_1(2', \bar{4}; \bar{3}, \bar{5}) \, t_1(\bar{6}, \bar{3}; 1, 2) \right.  \nonumber \\
&\hs*{5mm} \left. + t_1(1', 2'; \bar{6}, \bar{4}) \, t_1(\bar{5}, \bar{4}; \bar{3}, 2) \, t_1(\bar{6}, \bar{3}; 1, \bar{5}) \right. \nonumber \\
&\hs*{5mm} \left. + \left(t_1(2', \bar{5}; \bar{6}, \bar{4}) \, t_1(\bar{3}, \bar{4}; 2, \bar{5}) \, t_1(1', \bar{6} ; 1, \bar{3}) - \left(1' \leftrightarrow 2' \right) 
\right)  \right.  \nonumber \\
&\hs*{5mm} \left. + \left(t_1(1', \bar{5}; \bar{6}, \bar{4}) \, t_1(\bar{3}, \bar{4}; 1, \bar{5}) \, t_1(2', \bar{6}; 2, \bar{3})   - \left(1' \leftrightarrow 2' \right) \right)   \right. \nonumber \\
&\hs*{5mm} \left. - \left( t_1(2', \bar{5}; 2, \bar{4}) \, t_1(\bar{6}, \bar{4}, \bar{3}, \bar{5})
\, t_1(1', \bar{3}; 1, \bar{6})  - \left(1' \leftrightarrow 2' \right) \right) \right] \nonumber \\
&\hs*{5mm} \times S(\bar{3}) \, S(\bar{4}) \, S(\bar{5}) \, S(\bar{6}) \,.    
\label{t3nf}
\end{align} 
This form is used in Eq.~(\ref{se4}) of the main text to obtain the self energy to fourth order in $t_1$.
By using the form of $t_2$, given in Eq.~(\ref{t2nf}), and the antisymmetry of $t_1$, it is easy to see
that the counter term, given in the second line of Eq.~(\ref{se4}), cancels against the term which arises
from the second line of Eq.~(\ref{t3nf}). This cancellation is physically necessary, because product terms like
$t_1(i, j ; \bar{m}, \bar{n}) \, t_1(\bar{m}, \bar{n}; k, l)$
would double count the contribution of ladder graphs. We then obtain finally 
\begin{align}
&\Sigma_4(1) = \left[ \frac{1}{2} \, t_1(1, \bar{2}; \bar{3}, \bar{4}) \, t_1(\bar{3}, \bar{7}; \bar{8}, \bar{6}) \,
t_1(\bar{4}, \bar{6} ; \bar{5}, \bar{7}) \, t_1(\bar{8}, \bar{5} ; 1, \bar{2}) \right.  \nonumber \\
& \left. + t_1(1, \bar{2}; \bar{3}, \bar{4}) \, t_1(\bar{4}, \bar{7}; \bar{8}, \bar{6}) \,
t_1(\bar{5}, \bar{6} ; \bar{2}, \bar{7}) \, t_1(\bar{3}, \bar{8} ; 1, \bar{5}) \right.   \nonumber \\
& \left. + t_1(1, \bar{2}; \bar{3}, \bar{4}) \, t_1(\bar{3}, \bar{7}; \bar{8}, \bar{6}) \,
t_1(\bar{5}, \bar{6} ;1, \bar{7}) \, t_1(\bar{4}, \bar{8} ; \bar{2}, \bar{5})  \right.  \nonumber \\
& \left. - t_1(1, \bar{2}; \bar{3}, \bar{4}) \, t_1(\bar{4}, \bar{7}; \bar{2}, \bar{6}) \,
t_1(\bar{8}, \bar{6} ; \bar{5}, \bar{7}) \, t_1(\bar{3}, \bar{5} ; 1, \bar{8}) \right] \nonumber \\
& \hs*{3mm} \times S(\bar{2}) \,  S(\bar{3}) \, S(\bar{4}) \,  S(\bar{5}) \, S(\bar{6}) \,  S(\bar{7}) \, S(\bar{8})\,.
\label{se4f}
\end{align}

%===============================================================================
%===============================================================================
%\bibliographystyle{apsrev4-1}
%\bibliography{bib}


\end{document}